\documentclass[letter]{aa}     

\usepackage{graphicx}
\usepackage{txfonts}
\usepackage{lipsum}
\usepackage{subcaption}       
\usepackage{lscape} 
\usepackage{placeins}           

\usepackage{color}
\usepackage{soul}

\usepackage{stfloats}

\begin{document}

   \title{Accretion Disk Sizes and Temperature Profiles in Lensed Quasars: NIR Microlensing Challenges Thin Disk Theory}

   \author{V. Motta\inst{1}
        \and E. Mediavilla\inst{2} 
        }

   \institute{Instituto de F\'{\i}sica y Astronom\'{\i}a, Universidad de Valpara\'{\i}so, Avda. Gran Breta\~na 1111, Valpara\'{\i}so, Chile\\
             \email{veronica.motta@uv.cl}
             \and Instituto de Astrof\'{\i}sica de Canarias, Tenerife, Spain\\ }

   \date{Received July 1st, 2026; Acepted August 8th, 2026}

  \abstract
     {Microlensing and reverberation mapping measurements of quasar accretion disk sizes and temperature gradients disagree with thin disk theory predictions. Previous microlensing results rely on heterogeneous wavelength coverage—primarily UV broad emission lines (BELs)  from small samples —probing the disk only out to a typical radius of $\lesssim$5 light days on average. }
   {We use microlensing estimates from an homogeneous sample of near-infrared (NIR) observations of lensed quasars (21 image-pairs from 7 lens systems) to extend disk size measurements out to 14 light days. This analysis leverages narrow emission lines (NELs), which provide a more reliable microlensing-free baseline than BEL cores.} 
   {We derive Bayesian accretion disk size estimates that reproduce the observed microlensing magnifications, as simulated from magnification maps.}
   {NEL-based sizes yield a logarithmic slope of $p=0.68\pm0.23$, consistent with prior estimates corresponding to inner disk regions ($r \lesssim$5 light days). }
   {Using a new homogeneous NIR dataset that allows us to reach radial distances of up to 14 light days, we find that accretion disks in these previously unexplored regions are also larger and exhibit steeper temperature gradients than thin disk theory predicts. The increased precision allows us to reject the theoretical logarithmic slope $p=4/3$ at the 98\% confidence level. Any hypothesis invoking BLR contamination to explain this discrepancy must account for how such contamination modulates the underlying accretion disk such that the combination of both results in a power law with logarithmic slope $p=0.68\pm0.23$ across a broad wavelength baseline spanning from $\sim$X-Ray to $\sim 5000$\AA.
   }

   \keywords{Gravitational lensing: micro --
                Gravitational lensing: strong -- galaxies: active -- accretion disks
               }
\titlerunning{AGN accretion disk from chromatic microlensing}
   \maketitle

\nolinenumbers

\section{Introduction}

It is well known that the accretion disks of distant active galactic nuclei (AGNs) are too small to be directly resolved with existing technology. Thus, our knowledge of their structure and physical conditions is obtained through indirect approaches. Variability studies \cite[reverberation mapping, RM;][]{bahcall1972,blandford1982}  and gravitational microlensing \citep[e.g.,][]{pooley2007,morgan2010,black2011, jjv2014,jjv2015}, are effective methods for exploring those regions.

The thin accretion disk model \citep{ss1973}, is the standard model to understand energy generation of AGNs. This model provides predictions for observable features such as the size of the emitting disk and how it varies with wavelength. Specifically, the relationship between disk size and wavelength, or the temperature profile of the disk, is a crucial test for this model. Far from the disk's inner edge, a temperature profile of $T \propto r^{-1/p}$ \citep{peterson1997} means a size $r_{\lambda} \propto \lambda^p$, where the characteristic size is defined as the radius at which the rest wavelength matches the disk temperature. In the case of the thin disk model, the temperature profile follows $T \propto r^{-3/4}$, resulting in $p=4/3$.

Disk sizes estimated from continuum RM observations have yield values systematically larger \citep[from 2 to 7 times, with average $\sim4$,][]{edelson2015,fausnaugh2016,jiang2017,mudd2018,homa2019,yu2020,guo2022,jha2022,gonzalez2025,son2025,thorn2025,miller2026} than those predicted from the standard thin disk theory. 

On the other hand, a highly effective method for determining the sizes of accretion disks is through gravitational microlensing of lensed quasars\footnote{Notice that RM applies mainly to AGNs of lower luminosity than quasars, because variability is faster in these objects, while single epoch microlensing is better suited to study quasars.} \citep{wam2006}. Although many of the microlensing studies are based on individual objects  
\citep{irwin1989,corrigan1991,wyithe2000,wozniak2000,yonehara2001,shalyapin2002,goicoechea2003,eigen2008,floyd2009,dai2010,morgan2010,poin2010,med2011b,jjv2012,jjv2014,black2015,motta2017,fian2018,cornachione2020,cornachione2020b,cornachione2020c,rojas2020,fian2021,fores2024,sorgen2025}  there are a few based on samples of lenses \cite[e.g.][]{pooley2007,morgan2010,black2011,jjv2014,jjv2015,cornachione2020c}.  These studies show that accretion disk sizes increase with wavelength, reporting a disk size that is a factor of $\sim 4$ larger than predicted  and a value for $p$ closer to 1 instead of $4/3$.

One way to resolve these inconsistencies between the standard thin disk model and the estimates either from RM or microlensing is to modify the geometry of the accretion disk adding a contaminating component. Among some of those modifications, we can mention the inclusion of disk wind \citep{sun2019,netzer2025,wang2025}, a rippled accretion disk \citep{starkey2023}, or an important contribution from the recombination continuum of the BLR \citep[e.g.][]{mchardy2018}. New microlensing based measurements of increased accuracy, covering a larger wavelength range can constrain these an other alternative explanations.

A crucial step to obtain the microlensing  magnification amplitude is the precise determination of the no-microlensing  baseline. This could be established, for instance, from the mean light curve (once is shifted by time delay), the magnitude in IR bands, the macro-model magnifications or the  emission lines. However, there are uncertainties related to each technique. Although broad-band light curves allow the correction for time delay, they might be affected by contamination of emission lines, dust extinction  \citep{falco1999} and microlensing \cite[see][for a recent review]{ver2024}.  In addition,  the use of light curves is an observationally expensive technique, difficult to apply to a large sample of systems. Macro-model magnifications could lead to large uncertainties (see, e.g., Mediavilla et al. 2025) and do not prevent the effects of extinction. Intrinsic flux ratios can be inferred from radio or IR measurements but they can be different to those of the continuum because of source size effects.  

The core of the broad emission lines (BEL), arising from a region large enough as to be insensitive to microlensing, is often used to establish the zero microlensing baseline, with the automatic cancellation of extinction effects when using an adjacent continuum to measure microlensing amplitude. Although the use of narrow emission lines (NEL), is even simpler as the whole line is supposed to arise from an extended region and, consequently,  no recipe is needed to define the line core, they are scarcely used because typical strong lines of the optical spectra of quasars show the NEL lines blended with BEL.

Recent HST observations allow us to overcome this problem by providing spectroscopic NIR observations of lensed quasars \citep{nierenberg2020}. The main objective of this work  is, then, to use these data including strong, non-blended narrow emission lines like [OIII]$\lambda \lambda$4960,5007 to access, via microlensing measurements, unexplored regions of the accretion disk. We will combine the NIR data with previous optical results to estimate the logarithmic slope of the radial temperature profile with the benefit of a significantly larger wavelength range.

\section{Data}

\cite{nierenberg2020} provide an homogeneous sample of quadruply lensed quasars observed with the HST/WFC3\footnote{programs GO-13732 and GO-15177} with the filter F140W and grisms G102 or G141 to obtain [NeIII]$\lambda \lambda$3870,3969 and [OIII]$\lambda \lambda$4960,5007 narrow emission lines\footnote{The spectra are shown in appendix A of \cite{nierenberg2020}.}. The authors model the continuum in the observed region as a straight-line, the emission lines with narrow and broad components if required, adding the broad FeII lines in the H$\beta$ region. To account for possible variations in the continuum slope and broad emission line width, \cite{nierenberg2020} allow for variations in the continuum slope and normalization, as well as centroid, width and amplitude of emission lines,  between each quasar image. We use their published flux ratios for the continuum and narrow emission lines to obtain the microlensing magnification amplitude $\Delta m ^{obs}$ (see Table \ref{tab:dm} and Section \ref{sec:method}).

Other relevant information, such as redshifts for the source and lens as well as convergence and shear at the position of the images, are obtained from the literature (see Table \ref{tab:kg}). Notice that for $\rm WFI2026-4536$ we use the lens redshift estimated by \cite{cornachione2020} as it is the one compatible with the time delay estimated for the system. The authors also provide $\kappa, \gamma$ for range of mass-to-light fraction in the lens galaxy and we selected the fraction producing 10\% of stellar mass to be consistent with the treatment of the other systems.

\section{Method} \label{sec:method}

Single epoch spectroscopy allow us to separate the continuum emission (prone to microlensing magnification) from the NEL \citep[assumed unaffected by microlensing,][]{med2009}. 
As reference image, we select the one with less theoretical magnification unless its a saddle point or is suspected to have dust extinction or microlensing ($A$ image for $\rm DESJ0405-3308$, $\rm PSJ1606-2333$, and $\rm DESJ2038-4008$; $B$ for $\rm WFI2026-4536$, and $\rm WFI2033-4723$; $C$ for $\rm SDSSJ1330+1810$; $D$ for $\rm RXJ0911+0551$).
The microlensing magnification amplitude is obtained as the magnitude difference between the continuum and the emission line for each image with respect to the the reference one: 
\begin{equation}
\Delta m_i = (m_i - m_{ref})_{micro} = (m_i - m_{ref})_{cont} - (m_i - m_{ref})_{NEL},
\end{equation}
where $i=1,2,3$, $(m_i - m_{ref})_{NEL}$ is the magnitude difference in the NEL and $(m_i - m_{ref})_{cont}$ is the magnitude difference in its underlying continuum. 

To estimate the average size of the accretion disk at a given reference wavelength,  we follow \cite{jjv2012} procedure convolving magnification maps with disks of different size (see details about the magnification maps in appendix \ref{sec:maps}).
The radial structure of the disk is described as a Gaussian profile $I(R) \propto \exp(-R^2/2r_s^2)$, where $r_s$ is the characteristic size and it is connected to the half-light radius by $R_{1/2} = 1.18r_s$. This size is related to the wavelength as $r_s(\lambda) \propto \lambda^p$, with $p = 4/3$ for the \cite{ss1973} model. 
The reference wavelength is taken to be the redest wavelength for which we have a measured microlensing magnification, which is 5007 \AA \, (restframe [OIII]).  
We use a natural logarithmic grid in $r_s$ such that $\ln r_{s} = 0.0, \dots, 3.22$ with steps of $0.05$.

Each Gaussian profile with value $r_{s}$ is convolved with the magnification maps, normalized to the mean, to obtain the microlensing magnification histogram ($N_{ij}(\Delta m_i)$) 
which is compared with the observed microlensing signal ($\Delta m ^{obs}_i)$. For a given $\ln (r_{s})$, the probability of obtaining $\Delta m^{obs}_i$ for the lens $l$ is calculated as  
\begin{equation}
    P_l(\Delta m^{obs}| \ln(r_{s})) \propto \sum_{i,j} N_{ij} e^{-\chi_i^2/2}, \hspace{0.5cm} \chi^2_i =\frac{(\Delta m^{obs}_i - \Delta m_i)^2}{\sigma^2},
    \label{eq_prob}
\end{equation}
where $N_{i,j}$ is the number of trials with $\Delta m_i$, and 
$\sigma$ is the error in $\Delta m^{obs}_i$ (see Table \ref{tab:dm}).  The joint likelihood for $\ln(r_s)$ is obtained by multiplying the individual $P_l$ for the  three image-pairs of the seven lenses:
\begin{equation}
    P(\Delta m^{obs}| \ln(r_{s}) ) \propto \prod_{l=1}^{3\times7} P_l(\Delta m^{obs}| \ln(r_{s}))
    \label{eq_prod}
\end{equation}
Using the thin disk relation $r_s(\lambda) \propto \lambda^p$, we compare our estimate for $r_s$ at 5007\AA,  with the one at 1026\AA, to obtain the temperature profile parameter as 
\begin{equation}
\label{eq:comb}
p = \frac{\ln[r_s(1026)/r_s(5007)]}{\ln(1026/5007)}.
\end{equation}

\section{Results}

The values for $\Delta m^{obs}_i$ and their estimated errors for each of  the 21 image-pairs from the seven lenses are shown in Table \ref{tab:dm}.  
To place the $\rm [NeIII]$ and $\rm [OIII]$ based measurements on a common NIR reference size, we apply an iterative rescaling procedure using $r_s \propto \lambda^p$ to convert the two $\rm [NeIII]$ measurements to the $\rm [OIII]$ wavelength (5007~\AA). In the first iteration, we adopt $p = 4/3$. For the subsequent iterations, we use the value of $p$ obtained in the previous step, which is derived by combining our estimated size with that measured by \cite{jjv2014} at 1026~\AA\ in Equation~(\ref{eq:comb}). The procedure is repeated until convergence is reached (see Table~\ref{tab:pvalues}).

Figure \ref{fig:jointpdf} shows the joint likelihood of the probabilities derived from the 21 image-pairs, ($P=\prod_{l=1}^{3\times7} P_l$), for the final iteration.  
According to this Figure and Table \ref{tab:pvalues},  the agreement between realizations is excellent.  
We obtain the size at 5007\AA, $\langle r_s \rangle= 13.3^{+2.9}_{-2.7} \, \rm light-days$ and the scaling parameter, $\langle p \rangle=0.68\pm0.23$. 

In the Appendix (see Figures \ref{fig:pdfs} and \ref{fig:pdfs2}) we show the single probabilities  
for each pair of images (left panels) and their conflation (right panels) for each one of the seven lens systems 
of the final iteration. The mean values of the PDFs and theirs $1\sigma$ dispersions are used as estimators of $r_s$  and its uncertainty (at each iteration, see Table \ref{tab:pvalues}). The final values of $r_s$ for each individual system are presented in Table \ref{tab:rsvalues}. The average of the individual values is $r_s=13.4_{-2.8}^{+2.9} \, \rm light-days$ in very good agreement with the estimate obtained from the conflation of the PDFs.

\begin{figure}
        \centering      
         \includegraphics[width=9cm]{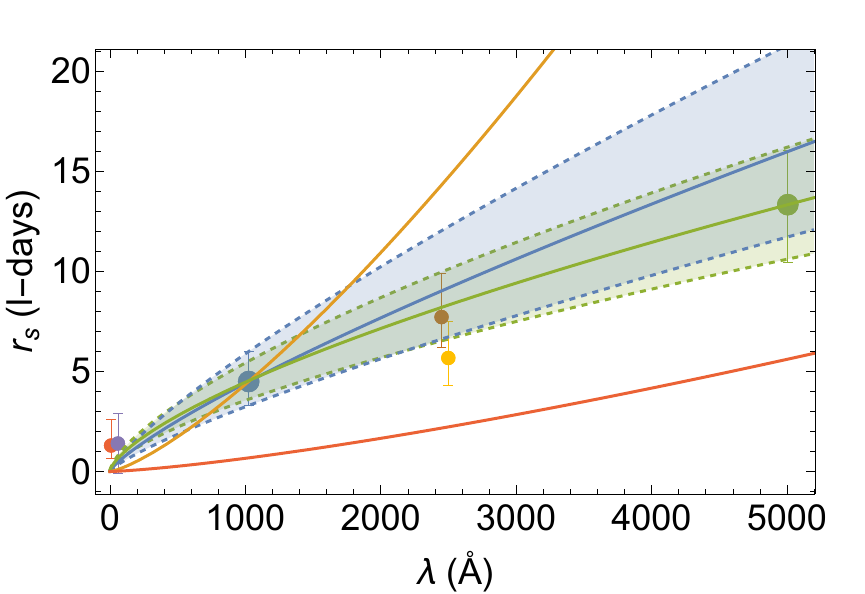}
      \caption{Size-scaling power-law estimated from $r_s$ measurements at 1026\AA\ (blue point) and 5007\AA\ (green point), for the average final iteration. Solid curves represent:  (red) thin-disk model for the average lensed quasar (\cite{avila2026}),  (orange) thin-disk model scaled to the measurement at 1026\AA, (blue) \cite{jjv2014} estimate,  and (green) this work. Blue/green dashed lines represent one sigma uncertainties. X-ray estimations from \cite{jjv2015} and \cite{pooley2007} are plotted at $\sim 10 \AA$, estimations for $10^9  M_{\odot}$ from \cite{morgan2010} (yellow) and \cite{black2011} (brown) are plotted at $\sim 2500 \AA$.} 
        \label{fig:p}
    \end{figure}

\section{Discussion and conclusions}

 In Figure \ref{fig:p} we show our size estimate inferred from the NIR observations at $\lambda5007$\AA, the size estimate from \cite{jjv2014} at $\lambda 1028$\AA\, and the $\lambda^p$ power-law curve ($p=0.68$) obtained using Equation (5) to combine these two measurements. Two additional size estimates corresponding to X-ray measurements from \cite{jjv2015} and \cite{pooley2007} are also plotted. Finally, we include the canonical thin-disk power-law curve, $\lambda^{4/3}$, normalized to the thin-disk size prediction at $\lambda 2500$\AA\ \citep{avila2026} and this same  curve normalized to the microlensing based size measurement at $\lambda 1028$\AA\ \citep{jjv2014}. Two direct consequences of Figure \ref{fig:p} are: (i) microlensing based measurements of the disk size are significantly larger than thin-disk predictions from X-ray to 5000\AA\ and (ii) the disk size scaling with wavelength, i.e. the  disk temperature profile, neither matches the theoretical thin-disk profile.

To assess the robustness and reach of these conclusions, we are going to discuss the reliability of the data and hypothesis used. In first place, the size predictions of the thin disk model are quite small ($<1$ light-day at $\lambda 1028$\AA) as compared with the Einstein ring radius of a star of $1M_\odot$. Accordingly, high microlensing magnifications would be expected. However, the statistics of microlensing magnification obtained with different observational techniques (lensed quasar photometric monitoring and single epoch spectroscopy) shows an scarcity of measured high microlensing magnifications,  $\Delta m< -0.6$ \cite[see, e.g.,][]{med2025}. Thus,  not only the present NIR measurements but also the generalized observational absence of high microlensing magnifications supports the consequence (i).

In principle, by virtue of the mass-size degeneracy of microlensing by point mass objects, considering stars of smaller mass would downsize the accretion discs. In fact, for stars  of $0.2M_\odot$ our estimate for the size at  $\lambda5007$\AA\  would match the thin disk prediction, although a stellar population of average mass $\lesssim0.1M_\odot$ would be required to do the same at $\lambda 1028$\AA. However, this degeneracy inherent to microlensing based methods is broken by reverberation mapping\footnote{A caveat is that continuum reverberation mapping may overestimate source sizes if variability propagates across the accretion disk more slowly than light.} (insensitive to stellar mass), which also measure larger sizes (by a factor $\sim$4 to 6) than thin-disk predictions \citep{edelson2015,jiang2017,homa2019,miller2026}\footnote{In any case, some  mitigation of the difference between microlensing based and theoretical sizes may not be dismissed out taking into account experimental uncertainties and the lack of knowledge about the mass of the stellar populations. In fact, the microlensing based estimate of the mean mass of the stellar population is $0.17M_\odot$ \citep{jjv2019}, which makes this possibility rather interesting.}.

Regarding the disk size scaling with wavelength, it is not affected by the mass of the microlenses at all. It may be affected by the fraction of mass in stars but, as we have commented in section 3, we have conservatively adopted a relatively low value. Using a larger one would result in a larger size and, consequently, in an even larger departure from the thin-disk model.
On the other hand, our  estimate of the scaling parameter, $\langle p \rangle=0.68\pm0.23$, is in agreement  with those calculated from reverberation mapping $p=0.92^{+0.61}_{-0.57}$ at 2500\AA,  \citep{guo2022} and   $p=0.78^{+0.41}_{-0.39}$, \citep{homa2019}, although their uncertainties are quite large and do not dismiss the thin-disk canonical value. Thanks to the better accuracy of our estimate, we can reject the theoretical value of $p= 4/3$ at the 98\% confidence level ($2.33\sigma$).

Several studies \citep{mchardy2018,sun2019,fian2023,starkey2023,netzer2025,wang2025} have suggested that the discrepancy between microlensing-based size estimates and the predictions of the standard thin disk model could be explained by the presence of an additional, more extended emission component — such as disk winds or the outer broad line region — that contributes a non-negligible fraction of the continuum flux and thus biases the inferred source size towards larger values. However, a key question is whether such an additional component can simultaneously account for the observed microlensing size excess while remaining consistent with all other observational constraints, including the integrated spectral energy distribution and the variability properties of the source. Critically, after the results presented here, any viable model must satisfy these conditions across the full wavelength range from X-rays to the optical at $5000\AA$ \, (matching the resulting $1/p=1/0.68$ observed temperature profile), since a component with the appropriate size and flux fraction at one wavelength must not overproduce or underproduce flux — nor predict inconsistent variability amplitudes — at other wavelengths, where independent constraints exist. This broad wavelength lever arm is, likely, an important limitation to any model devised to reconcile the standard thin-disk model with the observations.

We can summarize the present study, based on a sample of 21 quasar lensed image-pairs, in the following conclusions:

1 - Using NIR emission lines to define the no-microlensing baseline, we have significantly extended the study of the quasar accretion disk size and its scaling with wavelength out to 14 light-days, a region three times larger than that previously explored by microlensing-based studies.

2 - We find that the size of this region is larger by a factor of three than the thin-disk prediction.

3 - We fit a power-law $\lambda^p$ to the microlensing based estimates at $\lambda 1028$\AA\ and $\lambda 5007$\AA, obtaining $\langle p \rangle=0.68\pm0.23$. Thus, we can reject the theoretical value of $p= 4/3$ at the 98\% confidence level.

4 - The consistency of these microlensing-based results from X-ray to NIR, and their overall agreement with the results from reverberation mapping in the optical, makes it more difficult to invoke contamination of the disk emission by a more extended geometrical component (a BLR wind, for instance) as a means of reconciling the discrepancy with the thin-disk model, since any such component would need to account for the energy balance across this entire wavelength range while simultaneously reproducing the observed wavelength dependence of size.

\begin{acknowledgements}
We are grateful for helpful comments provided by the anonymous referee.  V.M. acknowledges support from ANID FONDECYT Regular grant number 1231418, and Centro de Astrof\'{\i}sica de Valpara\'{\i}so CIDI 21. E.M. acknowledges support from grant PID2025-160091NB-C31, financed by the Spanish Ministerio de Ciencia, Innovaci\'on y Universidades/Agencia Estatal de Investigaci\'on.
\end{acknowledgements}

\bibliographystyle{aa} 
\bibliography{ms.bib}

@ARTICLE{agnello2018a,
       author = {{Agnello}, A. and {Lin}, H. and {Kuropatkin}, N. and {Buckley-Geer}, E. and {Anguita}, T. and {Schechter}, P.~L. and {Morishita}, T. and {Motta}, V. and {Rojas}, K. and {Treu}, T. and {Amara}, A. and {Auger}, M.~W. and {Courbin}, F. and {Fassnacht}, C.~D. and {Frieman}, J. and {More}, A. and {Marshall}, P.~J. and {McMahon}, R.~G. and {Meylan}, G. and {Suyu}, S.~H. and {Glazebrook}, K. and {Morgan}, N. and {Nord}, B. and {Abbott}, T.~M.~C. and {Abdalla}, F.~B. and {Annis}, J. and {Bechtol}, K. and {Benoit-L{\'e}vy}, A. and {Bertin}, E. and {Bernstein}, R.~A. and {Brooks}, D. and {Burke}, D.~L. and {Rosell}, A. Carnero and {Carretero}, J. and {Cunha}, C.~E. and {D'Andrea}, C.~B. and {da Costa}, L.~N. and {Desai}, S. and {Drlica-Wagner}, A. and {Eifler}, T.~F. and {Flaugher}, B. and {Garc{\'\i}a-Bellido}, J. and {Gaztanaga}, E. and {Gerdes}, D.~W. and {Gruen}, D. and {Gruendl}, R.~A. and {Gschwend}, J. and {Gutierrez}, G. and {Honscheid}, K. and {James}, D.~J. and {Kuehn}, K. and {Lahav}, O. and {Lima}, M. and {Maia}, M.~A.~G. and {March}, M. and {Menanteau}, F. and {Miquel}, R. and {Ogando}, R.~L.~C. and {Plazas}, A.~A. and {Sanchez}, E. and {Scarpine}, V. and {Schindler}, R. and {Schubnell}, M. and {Sevilla-Noarbe}, I. and {Smith}, M. and {Soares-Santos}, M. and {Sobreira}, F. and {Suchyta}, E. and {Swanson}, M.~E.~C. and {Tarle}, G. and {Tucker}, D. and {Wechsler}, R.},
        title = "{DES meets Gaia: discovery of strongly lensed quasars from a multiplet search}",
      journal = {\mnras},
         year = 2018,
        month = oct,
       volume = {479},
       number = {4},
        pages = {4345-4354},
          doi = {10.1093/mnras/sty1419},
archivePrefix = {arXiv},
       eprint = {1711.03971},
 primaryClass = {astro-ph.CO},
       adsurl = {https://ui.adsabs.harvard.edu/abs/2018MNRAS.479.4345A}
}

@ARTICLE{anguita2018,
       author = {{Anguita}, T. and {Schechter}, P.~L. and {Kuropatkin}, N. and {Morgan}, N.~D. and {Ostrovski}, F. and {Abramson}, L.~E. and {Agnello}, A. and {Apostolovski}, Y. and {Fassnacht}, C.~D. and {Hsueh}, J.~W. and {Motta}, V. and {Rojas}, K. and {Rusu}, C.~E. and {Treu}, T. and {Williams}, P. and {Auger}, M. and {Buckley-Geer}, E. and {Lin}, H. and {McMahon}, R. and {Abbott}, T.~M.~C. and {Allam}, S. and {Annis}, J. and {Bernstein}, R.~A. and {Bertin}, E. and {Brooks}, D. and {Burke}, D.~L. and {Carnero Rosell}, A. and {Carrasco-Kind}, M. and {Carretero}, J. and {Cunha}, C.~E. and {D'Andrea}, C.~B. and {De Vicente}, J. and {DePoy}, D.~L. and {Desai}, S. and {Diehl}, H.~T. and {Doel}, P. and {Flaugher}, B. and {Garc{\'\i}a-Bellido}, J. and {Gerdes}, D.~W. and {Gruen}, D. and {Gruendl}, R.~A. and {Gschwend}, J. and {Hartley}, W.~G. and {Hollowood}, D.~L. and {Honscheid}, K. and {James}, D.~J. and {Kuehn}, K. and {Lima}, M. and {Maia}, M.~A.~G. and {Miquel}, R. and {Plazas}, A.~A. and {Sanchez}, E. and {Scarpine}, V. and {Smith}, M. and {Soares-Santos}, M. and {Sobreira}, F. and {Suchyta}, E. and {Tarle}, G. and {Walker}, A.~R.},
        title = "{The STRong lensing Insights into the Dark Energy Survey (STRIDES) 2016 follow-up campaign - II. New quasar lenses from double component fitting}",
      journal = {\mnras},
         year = 2018,
        month = nov,
       volume = {480},
       number = {4},
        pages = {5017-5028},
          doi = {10.1093/mnras/sty2172},
archivePrefix = {arXiv},
       eprint = {1805.12151},
 primaryClass = {astro-ph.GA},
       adsurl = {https://ui.adsabs.harvard.edu/abs/2018MNRAS.480.5017A}
}

@ARTICLE{avila2026,
	author = {{{\'A}vila-Vera}, F. and {Motta}, V. and {Mediavilla}, E.},
	title = "{Microlensing timescales and flux magnification probabilities of a sample of 204 lensed quasars}",
	journal = {\aap},
	year = 2026,
	month = jul,
	volume = {711},
	eid = {A123},
	pages = {A123},
	doi = {10.1051/0004-6361/202557467},
	archivePrefix = {arXiv},
	eprint = {2605.03905},
	primaryClass = {astro-ph.GA},
	adsurl = {https://ui.adsabs.harvard.edu/abs/2026A&A...711A.123A}
}

@ARTICLE{bade1997,
       author = {{Bade}, N. and {Siebert}, J. and {Lopez}, S. and {Voges}, W. and {Reimers}, D.},
        title = "{RX J0911.4+0551: A new multiple QSO selected from the ROSAT All-Sky Survey.}",
      journal = {\aap},
         year = 1997,
        month = jan,
       volume = {317},
        pages = {L13-L16},
       adsurl = {https://ui.adsabs.harvard.edu/abs/1997A&A...317L..13B}
}

@ARTICLE{bahcall1972,
       author = {{Bahcall}, John N. and {Kozlovsky}, Ben-Zion and {Salpeter}, E.~E.},
        title = "{On the Time Dependence of Emission-Line Strengths from a Photoionized Nebula}",
      journal = {\apj},
         year = 1972,
        month = feb,
       volume = {171},
        pages = {467},
          doi = {10.1086/151300},
       adsurl = {https://ui.adsabs.harvard.edu/abs/1972ApJ...171..467B}
}

@ARTICLE{black2015,
       author = {{Blackburne}, Jeffrey A. and {Kochanek}, Christopher S. and {Chen}, Bin and {Dai}, Xinyu and {Chartas}, George},
        title = "{The Structure of HE 1104-1805 from Infrared to X-Ray}",
      journal = {\apj},
         year = 2015,
        month = jan,
       volume = {798},
       number = {2},
          eid = {95},
        pages = {95},
          doi = {10.1088/0004-637X/798/2/95},
archivePrefix = {arXiv},
       eprint = {1304.1620},
 primaryClass = {astro-ph.CO},
       adsurl = {https://ui.adsabs.harvard.edu/abs/2015ApJ...798...95B}
}

@ARTICLE{black2011,
       author = {{Blackburne}, Jeffrey A. and {Pooley}, David and {Rappaport}, Saul and {Schechter}, Paul L.},
        title = "{Sizes and Temperature Profiles of Quasar Accretion Disks from Chromatic Microlensing}",
      journal = {\apj},
         year = 2011,
        month = mar,
       volume = {729},
       number = {1},
          eid = {34},
        pages = {34},
          doi = {10.1088/0004-637X/729/1/34},
archivePrefix = {arXiv},
       eprint = {1007.1665},
 primaryClass = {astro-ph.CO},
       adsurl = {https://ui.adsabs.harvard.edu/abs/2011ApJ...729...34B}
}

@ARTICLE{blandford1982,
       author = {{Blandford}, R.~D. and {McKee}, C.~F.},
        title = "{Reverberation mapping of the emission line regions of Seyfert galaxies and quasars.}",
      journal = {\apj},
         year = 1982,
        month = apr,
       volume = {255},
        pages = {419-439},
          doi = {10.1086/159843},
       adsurl = {https://ui.adsabs.harvard.edu/abs/1982ApJ...255..419B}
}

@ARTICLE{cornachione2020,
       author = {{Cornachione}, Matthew A. and {Morgan}, Christopher W. and {Millon}, Martin and {Bentz}, Misty C. and {Courbin}, Frederic and {Bonvin}, Vivien and {Falco}, Emilio E.},
        title = "{A Microlensing Accretion Disk Size Measurement in the Lensed Quasar WFI 2026-4536}",
      journal = {\apj},
         year = 2020,
        month = jun,
       volume = {895},
       number = {2},
          eid = {125},
        pages = {125},
          doi = {10.3847/1538-4357/ab557a},
archivePrefix = {arXiv},
       eprint = {1911.06218},
 primaryClass = {astro-ph.HE},
       adsurl = {https://ui.adsabs.harvard.edu/abs/2020ApJ...895..125C}
}

@ARTICLE{cornachione2020b,
       author = {{Cornachione}, Matthew A. and {Morgan}, Christopher W. and {Burger}, Hayden R. and {Shalyapin}, Vyacheslav N. and {Goicoechea}, Luis J. and {Vrba}, Frederick J. and {Dahm}, Scott E. and {Tilleman}, Trudy M.},
        title = "{Near-infrared and Optical Continuum Emission Region Size Measurements in the Gravitationally lensed Quasars Q0957+561 and SBS0909+532}",
      journal = {\apj},
         year = 2020,
        month = dec,
       volume = {905},
       number = {1},
          eid = {7},
        pages = {7},
          doi = {10.3847/1538-4357/abc25d},
archivePrefix = {arXiv},
       eprint = {2012.05856},
 primaryClass = {astro-ph.GA},
       adsurl = {https://ui.adsabs.harvard.edu/abs/2020ApJ...905....7C}
}

@ARTICLE{cornachione2020c,
       author = {{Cornachione}, Matthew A. and {Morgan}, Christopher W.},
        title = "{Quasar Microlensing Variability Studies Favor Shallow Accretion Disk Temperature Profiles}",
      journal = {\apj},
         year = 2020,
        month = jun,
       volume = {895},
       number = {2},
          eid = {93},
        pages = {93},
          doi = {10.3847/1538-4357/ab8aed},
archivePrefix = {arXiv},
       eprint = {2006.07243},
 primaryClass = {astro-ph.HE},
       adsurl = {https://ui.adsabs.harvard.edu/abs/2020ApJ...895...93C}
}

@ARTICLE{corrigan1991,
       author = {{Corrigan}, R.~T. and {Irwin}, M.~J. and {Arnaud}, J. and {Fahlman}, G.~G. and {Fletcher}, J.~M. and {Hewett}, P.~C. and {Hewitt}, J.~N. and {Le Fevre}, O. and {McClure}, R. and {Pritchet}, C.~J. and {Schneider}, D.~P. and {Turner}, E.~L. and {Webster}, R.~L. and {Yee}, H.~K.~C.},
        title = "{Initial Light Curve of Q2237+0305}",
      journal = {\aj},
         year = 1991,
        month = jul,
       volume = {102},
        pages = {34},
          doi = {10.1086/115856},
       adsurl = {https://ui.adsabs.harvard.edu/abs/1991AJ....102...34C}
}

@ARTICLE{dai2010,
       author = {{Dai}, X. and {Kochanek}, C.~S. and {Chartas}, G. and {Koz{\l}owski}, S. and {Morgan}, C.~W. and {Garmire}, G. and {Agol}, E.},
        title = "{The Sizes of the X-ray and Optical Emission Regions of RXJ 1131-1231}",
      journal = {\apj},
         year = 2010,
        month = jan,
       volume = {709},
       number = {1},
        pages = {278-285},
          doi = {10.1088/0004-637X/709/1/278},
archivePrefix = {arXiv},
       eprint = {0906.4342},
 primaryClass = {astro-ph.HE},
       adsurl = {https://ui.adsabs.harvard.edu/abs/2010ApJ...709..278D}
}

@ARTICLE{edelson2015,
       author = {{Edelson}, R. and {Gelbord}, J.~M. and {Horne}, K. and {McHardy}, I.~M. and {Peterson}, B.~M. and {Ar{\'e}valo}, P. and {Breeveld}, A.~A. and {De Rosa}, G. and {Evans}, P.~A. and {Goad}, M.~R. and {Kriss}, G.~A. and {Brandt}, W.~N. and {Gehrels}, N. and {Grupe}, D. and {Kennea}, J.~A. and {Kochanek}, C.~S. and {Nousek}, J.~A. and {Papadakis}, I. and {Siegel}, M. and {Starkey}, D. and {Uttley}, P. and {Vaughan}, S. and {Young}, S. and {Barth}, A.~J. and {Bentz}, M.~C. and {Brewer}, B.~J. and {Crenshaw}, D.~M. and {Dalla Bont{\`a}}, E. and {De Lorenzo-C{\'a}ceres}, A. and {Denney}, K.~D. and {Dietrich}, M. and {Ely}, J. and {Fausnaugh}, M.~M. and {Grier}, C.~J. and {Hall}, P.~B. and {Kaastra}, J. and {Kelly}, B.~C. and {Korista}, K.~T. and {Lira}, P. and {Mathur}, S. and {Netzer}, H. and {Pancoast}, A. and {Pei}, L. and {Pogge}, R.~W. and {Schimoia}, J.~S. and {Treu}, T. and {Vestergaard}, M. and {Villforth}, C. and {Yan}, H. and {Zu}, Y.},
        title = "{Space Telescope and Optical Reverberation Mapping Project. II. Swift and HST Reverberation Mapping of the Accretion Disk of NGC 5548}",
      journal = {\apj},
         year = 2015,
        month = jun,
       volume = {806},
       number = {1},
          eid = {129},
        pages = {129},
          doi = {10.1088/0004-637X/806/1/129},
archivePrefix = {arXiv},
       eprint = {1501.05951},
 primaryClass = {astro-ph.GA},
       adsurl = {https://ui.adsabs.harvard.edu/abs/2015ApJ...806..129E}
}

@ARTICLE{eigen2008,
       author = {{Eigenbrod}, A. and {Courbin}, F. and {Meylan}, G. and {Agol}, E. and {Anguita}, T. and {Schmidt}, R.~W. and {Wambsganss}, J.},
        title = "{Microlensing variability in the gravitationally lensed quasar QSO 2237+0305 {\ensuremath{\equiv}} the Einstein Cross. II. Energy profile of the accretion disk}",
      journal = {\aap},
         year = 2008,
        month = nov,
       volume = {490},
       number = {3},
        pages = {933-943},
          doi = {10.1051/0004-6361:200810729},
archivePrefix = {arXiv},
       eprint = {0810.0011},
 primaryClass = {astro-ph},
       adsurl = {https://ui.adsabs.harvard.edu/abs/2008A&A...490..933E}
}

@ARTICLE{falco1999,
       author = {{Falco}, E.~E. and {Impey}, C.~D. and {Kochanek}, C.~S. and {Leh{\'a}r}, J. and {McLeod}, B.~A. and {Rix}, H.-W. and {Keeton}, C.~R. and {Mu{\~n}oz}, J.~A. and {Peng}, C.~Y.},
        title = "{Dust and Extinction Curves in Galaxies with z>0: The Interstellar Medium of Gravitational Lens Galaxies}",
      journal = {\apj},
         year = 1999,
        month = oct,
       volume = {523},
       number = {2},
        pages = {617-632},
          doi = {10.1086/307758},
archivePrefix = {arXiv},
       eprint = {astro-ph/9901037},
 primaryClass = {astro-ph},
       adsurl = {https://ui.adsabs.harvard.edu/abs/1999ApJ...523..617F}
}

@ARTICLE{fausnaugh2016,
       author = {{Fausnaugh}, M.~M. and {Denney}, K.~D. and {Barth}, A.~J. and {Bentz}, M.~C. and {Bottorff}, M.~C. and {Carini}, M.~T. and {Croxall}, K.~V. and {De Rosa}, G. and {Goad}, M.~R. and {Horne}, Keith and {Joner}, M.~D. and {Kaspi}, S. and {Kim}, M. and {Klimanov}, S.~A. and {Kochanek}, C.~S. and {Leonard}, D.~C. and {Netzer}, H. and {Peterson}, B.~M. and {Schn{\"u}lle}, K. and {Sergeev}, S.~G. and {Vestergaard}, M. and {Zheng}, W. -K. and {Zu}, Y. and {Anderson}, M.~D. and {Ar{\'e}valo}, P. and {Bazhaw}, C. and {Borman}, G.~A. and {Boroson}, T.~A. and {Brandt}, W.~N. and {Breeveld}, A.~A. and {Brewer}, B.~J. and {Cackett}, E.~M. and {Crenshaw}, D.~M. and {Dalla Bont{\`a}}, E. and {De Lorenzo-C{\'a}ceres}, A. and {Dietrich}, M. and {Edelson}, R. and {Efimova}, N.~V. and {Ely}, J. and {Evans}, P.~A. and {Filippenko}, A.~V. and {Flatland}, K. and {Gehrels}, N. and {Geier}, S. and {Gelbord}, J.~M. and {Gonzalez}, L. and {Gorjian}, V. and {Grier}, C.~J. and {Grupe}, D. and {Hall}, P.~B. and {Hicks}, S. and {Horenstein}, D. and {Hutchison}, T. and {Im}, M. and {Jensen}, J.~J. and {Jones}, J. and {Kaastra}, J. and {Kelly}, B.~C. and {Kennea}, J.~A. and {Kim}, S.~C. and {Korista}, K.~T. and {Kriss}, G.~A. and {Lee}, J.~C. and {Lira}, P. and {MacInnis}, F. and {Manne-Nicholas}, E.~R. and {Mathur}, S. and {McHardy}, I.~M. and {Montouri}, C. and {Musso}, R. and {Nazarov}, S.~V. and {Norris}, R.~P. and {Nousek}, J.~A. and {Okhmat}, D.~N. and {Pancoast}, A. and {Papadakis}, I. and {Parks}, J.~R. and {Pei}, L. and {Pogge}, R.~W. and {Pott}, J. -U. and {Rafter}, S.~E. and {Rix}, H. -W. and {Saylor}, D.~A. and {Schimoia}, J.~S. and {Siegel}, M. and {Spencer}, M. and {Starkey}, D. and {Sung}, H. -I. and {Teems}, K.~G. and {Treu}, T. and {Turner}, C.~S. and {Uttley}, P. and {Villforth}, C. and {Weiss}, Y. and {Woo}, J. -H. and {Yan}, H. and {Young}, S.},
        title = "{Space Telescope and Optical Reverberation Mapping Project. III. Optical Continuum Emission and Broadband Time Delays in NGC 5548}",
      journal = {\apj},
         year = 2016,
        month = apr,
       volume = {821},
       number = {1},
          eid = {56},
        pages = {56},
          doi = {10.3847/0004-637X/821/1/56},
archivePrefix = {arXiv},
       eprint = {1510.05648},
 primaryClass = {astro-ph.GA},
       adsurl = {https://ui.adsabs.harvard.edu/abs/2016ApJ...821...56F}
}

@ARTICLE{fian2018,
       author = {{Fian}, C. and {Mediavilla}, E. and {Jim{\'e}nez-Vicente}, J. and {Mu{\~n}oz}, J.~A. and {Hanslmeier}, A.},
        title = "{Estimate of the Accretion Disk Size in the Gravitationally Lensed Quasar HE 0435-1223 Using Microlensing Magnification Statistics}",
      journal = {\apj},
         year = 2018,
        month = dec,
       volume = {869},
       number = {2},
          eid = {132},
        pages = {132},
          doi = {10.3847/1538-4357/aaeed5},
archivePrefix = {arXiv},
       eprint = {1811.03312},
 primaryClass = {astro-ph.GA},
       adsurl = {https://ui.adsabs.harvard.edu/abs/2018ApJ...869..132F}
}

@ARTICLE{fian2021,
       author = {{Fian}, C. and {Mediavilla}, E. and {Jim{\'e}nez-Vicente}, J. and {Motta}, V. and {Mu{\~n}oz}, J.~A. and {Chelouche}, D. and {Gom{\'e}z-Alvarez}, P. and {Rojas}, K. and {Hanslmeier}, A.},
        title = "{Revealing the structure of the lensed quasar Q 0957+561. I. Accretion disk size}",
      journal = {\aap},
         year = 2021,
        month = oct,
       volume = {654},
          eid = {A70},
        pages = {A70},
          doi = {10.1051/0004-6361/202039854},
archivePrefix = {arXiv},
       eprint = {2108.05212},
 primaryClass = {astro-ph.GA},
       adsurl = {https://ui.adsabs.harvard.edu/abs/2021A&A...654A..70F}
}

@ARTICLE{fian2023,
       author = {{Fian}, C. and {Chelouche}, D. and {Kaspi}, S.},
        title = "{Diffuse emission in microlensed quasars and its implications for accretion-disk physics}",
      journal = {\aap},
         year = 2023,
        month = sep,
       volume = {677},
          eid = {A94},
        pages = {A94},
          doi = {10.1051/0004-6361/202346766},
archivePrefix = {arXiv},
       eprint = {2307.14824},
 primaryClass = {astro-ph.GA},
       adsurl = {https://ui.adsabs.harvard.edu/abs/2023A&A...677A..94F}
}

@ARTICLE{floyd2009,
       author = {{Floyd}, David J.~E. and {Bate}, N.~F. and {Webster}, R.~L.},
        title = "{The accretion disc in the quasar SDSS J0924+0219}",
      journal = {\mnras},
         year = 2009,
        month = sep,
       volume = {398},
       number = {1},
        pages = {233-239},
          doi = {10.1111/j.1365-2966.2009.15045.x},
archivePrefix = {arXiv},
       eprint = {0905.2651},
 primaryClass = {astro-ph.HE},
       adsurl = {https://ui.adsabs.harvard.edu/abs/2009MNRAS.398..233F}
}

@ARTICLE{fores2024,
       author = {{For{\'e}s-Toribio}, R. and {Mu{\~n}oz}, J.~A. and {Fian}, C. and {Jim{\'e}nez-Vicente}, J. and {Mediavilla}, E.},
        title = "{Microlensing analysis of 14.5-year light curves in SDSS J1004+4112: Quasar accretion disk size and intracluster stellar mass fraction}",
      journal = {\aap},
         year = 2024,
        month = nov,
       volume = {691},
          eid = {A97},
        pages = {A97},
          doi = {10.1051/0004-6361/202347378},
archivePrefix = {arXiv},
       eprint = {2410.06853},
 primaryClass = {astro-ph.GA},
       adsurl = {https://ui.adsabs.harvard.edu/abs/2024A&A...691A..97F}
}

@ARTICLE{goicoechea2003,
	author = {{Goicoechea}, L.~J. and {Alcalde}, D. and {Mediavilla}, E. and {Mu{\~n}oz}, J.~A.},
	title = "{Determination of the properties of the central engine in microlensed QSOs}",
	journal = {\aap},
	year = 2003,
	month = jan,
	volume = {397},
	pages = {517-525},
	doi = {10.1051/0004-6361:20021535},
	archivePrefix = {arXiv},
	eprint = {astro-ph/0210291},
	primaryClass = {astro-ph},
	adsurl = {https://ui.adsabs.harvard.edu/abs/2003A&A...397..517G}
}

@ARTICLE{gonzalez2025,
       author = {{Gonzalez-Buitrago}, D. and {Barth}, A.~J. and {Edelson}, R. and {Hern{\'a}ndez Santisteban}, J.~V. and {Horne}, Keith and {Schmidt}, T. and {Li}, Yan-Rong and {Guo}, Hengxiao and {Joner}, M.~D. and {Cackett}, E. and {Gelbord}, J. and {Bentz}, M.~C. and {Brandt}, W.~N. and {Goad}, M. and {Korista}, K. and {Vestergaard}, M. and {Villforth}, C. and {Breeveld}, A. and {Brink}, T.~G. and {Corsini}, E.~M. and {Dalla Bont{\`a}}, E. and {Ferland}, Gary J. and {Filippenko}, A.~V. and {Garc{\'\i}a-D{\'\i}az}, Ma T. and {Hallum}, M. and {Horst}, J.~C. and {Kim}, M. and {Krongold}, Y. and {Kruger}, J. and {Kuhn}, B. and {Kumar}, S. and {Mehdipour}, M. and {Morelli}, L. and {Mathur}, S. and {Netzer}, H. and {Ochner}, P. and {Pagotto}, I. and {Pizzella}, A. and {Sand}, D.~J. and {Siviero}, A. and {Spencer}, M. and {Sung}, H. and {Vaughan}, S. and {Winkler}, H. and {Zheng}, W.},
        title = "{Departures from standard disc predictions in intensive ground-based monitoring of three AGNs}",
      journal = {\mnras},
         year = 2025,
        month = sep,
       volume = {542},
       number = {3},
        pages = {2572-2596},
          doi = {10.1093/mnras/staf1334},
archivePrefix = {arXiv},
       eprint = {2508.08720},
 primaryClass = {astro-ph.GA},
       adsurl = {https://ui.adsabs.harvard.edu/abs/2025MNRAS.542.2572G}
}

@ARTICLE{guo2022,
       author = {{Guo}, Hengxiao and {Barth}, Aaron J. and {Wang}, Shu},
        title = "{Active Galactic Nuclei Continuum Reverberation Mapping Based on Zwicky Transient Facility Light Curves}",
      journal = {\apj},
         year = 2022,
        month = nov,
       volume = {940},
       number = {1},
          eid = {20},
        pages = {20},
          doi = {10.3847/1538-4357/ac96ec},
archivePrefix = {arXiv},
       eprint = {2207.06432},
 primaryClass = {astro-ph.GA},
       adsurl = {https://ui.adsabs.harvard.edu/abs/2022ApJ...940...20G}
}

@ARTICLE{homa2019,
       author = {{Homayouni}, Y. and {Trump}, Jonathan R. and {Grier}, C.~J. and {Shen}, Yue and {Starkey}, D.~A. and {Brandt}, W.~N. and {Fonseca Alvarez}, G. and {Hall}, P.~B. and {Horne}, Keith and {Kinemuchi}, Karen and {I-Hsiu Li}, Jennifer and {McGreer}, Ian D. and {Sun}, Mouyuan and {Ho}, L.~C. and {Schneider}, D.~P.},
        title = "{The Sloan Digital Sky Survey Reverberation Mapping Project: Accretion Disk Sizes from Continuum Lags}",
      journal = {\apj},
         year = 2019,
        month = aug,
       volume = {880},
       number = {2},
          eid = {126},
        pages = {126},
          doi = {10.3847/1538-4357/ab2638},
archivePrefix = {arXiv},
       eprint = {1806.08360},
 primaryClass = {astro-ph.GA},
       adsurl = {https://ui.adsabs.harvard.edu/abs/2019ApJ...880..126H}
}

@ARTICLE{irwin1989,
       author = {{Irwin}, M.~J. and {Webster}, R.~L. and {Hewett}, P.~C. and {Corrigan}, R.~T. and {Jedrzejewski}, R.~I.},
        title = "{Photometric Variations in the Q2237+0305 System: First Detection of a Microlensing Event}",
      journal = {\aj},
         year = 1989,
        month = dec,
       volume = {98},
        pages = {1989},
          doi = {10.1086/115272},
       adsurl = {https://ui.adsabs.harvard.edu/abs/1989AJ.....98.1989I}
}

@ARTICLE{jjv2012,
       author = {{Jim{\'e}nez-Vicente}, J. and {Mediavilla}, E. and {Mu{\~n}oz}, J.~A. and {Kochanek}, C.~S.},
        title = "{A Robust Determination of the Size of Quasar Accretion Disks Using Gravitational Microlensing}",
      journal = {\apj},
         year = 2012,
        month = jun,
       volume = {751},
       number = {2},
          eid = {106},
        pages = {106},
          doi = {10.1088/0004-637X/751/2/106},
archivePrefix = {arXiv},
       eprint = {1201.3187},
 primaryClass = {astro-ph.CO},
       adsurl = {https://ui.adsabs.harvard.edu/abs/2012ApJ...751..106J}
}

@ARTICLE{jjv2019,
       author = {{Jim{\'e}nez-Vicente}, J. and {Mediavilla}, E.},
        title = "{The Initial Mass Function of Lens Galaxies from Quasar Microlensing}",
      journal = {\apj},
         year = 2019,
        month = nov,
       volume = {885},
       number = {1},
          eid = {75},
        pages = {75},
          doi = {10.3847/1538-4357/ab46b8},
archivePrefix = {arXiv},
       eprint = {1910.10509},
 primaryClass = {astro-ph.GA},
       adsurl = {https://ui.adsabs.harvard.edu/abs/2019ApJ...885...75J}
}

@ARTICLE{jha2022,
       author = {{Jha}, Vivek Kumar and {Joshi}, Ravi and {Chand}, Hum and {Wu}, Xue-Bing and {Ho}, Luis C. and {Rastogi}, Shantanu and {Ma}, Qinchun},
        title = "{Accretion disc sizes from continuum reverberation mapping of AGN selected from the ZTF survey}",
      journal = {\mnras},
         year = 2022,
        month = apr,
       volume = {511},
       number = {2},
        pages = {3005-3016},
          doi = {10.1093/mnras/stac109},
archivePrefix = {arXiv},
       eprint = {2109.05036},
 primaryClass = {astro-ph.GA},
       adsurl = {https://ui.adsabs.harvard.edu/abs/2022MNRAS.511.3005J}
}

@ARTICLE{jiang2017,
       author = {{Jiang}, Yan-Fei and {Green}, Paul J. and {Greene}, Jenny E. and {Morganson}, Eric and {Shen}, Yue and {Pancoast}, Anna and {MacLeod}, Chelsea L. and {Anderson}, Scott F. and {Brandt}, W.~N. and {Grier}, C.~J. and {Rix}, H.-W. and {Ruan}, John J. and {Protopapas}, Pavlos and {Scott}, Caroline and {Burgett}, W.~S. and {Hodapp}, K.~W. and {Huber}, M.~E. and {Kaiser}, N. and {Kudritzki}, R.~P. and {Magnier}, E.~A. and {Metcalfe}, N. and {Tonry}, J.~T. and {Wainscoat}, R.~J. and {Waters}, C.},
        title = "{Detection of Time Lags between Quasar Continuum Emission Bands Based On Pan-STARRS Light Curves}",
      journal = {\apj},
         year = 2017,
        month = feb,
       volume = {836},
       number = {2},
          eid = {186},
        pages = {186},
          doi = {10.3847/1538-4357/aa5b91},
archivePrefix = {arXiv},
       eprint = {1612.08747},
 primaryClass = {astro-ph.HE},
       adsurl = {https://ui.adsabs.harvard.edu/abs/2017ApJ...836..186J}
}

@ARTICLE{jjv2014,
       author = {{Jim{\'e}nez-Vicente}, J. and {Mediavilla}, E. and {Kochanek}, C.~S. and {Mu{\~n}oz}, J.~A. and {Motta}, V. and {Falco}, E. and {Mosquera}, A.~M.},
        title = "{The Average Size and Temperature Profile of Quasar Accretion Disks}",
      journal = {\apj},
         year = 2014,
        month = mar,
       volume = {783},
       number = {1},
          eid = {47},
        pages = {47},
          doi = {10.1088/0004-637X/783/1/47},
archivePrefix = {arXiv},
       eprint = {1401.2785},
 primaryClass = {astro-ph.CO},
       adsurl = {https://ui.adsabs.harvard.edu/abs/2014ApJ...783...47J}
}

@ARTICLE{jjv2015,
       author = {{Jim{\'e}nez-Vicente}, J. and {Mediavilla}, E. and {Kochanek}, C.~S. and {Mu{\~n}oz}, J.~A.},
        title = "{Probing the Dark Matter Radial Profile in Lens Galaxies and the Size of X-Ray Emitting Region in Quasars with Microlensing}",
      journal = {\apj},
         year = 2015,
        month = jun,
       volume = {806},
       number = {2},
          eid = {251},
        pages = {251},
          doi = {10.1088/0004-637X/806/2/251},
archivePrefix = {arXiv},
       eprint = {1502.00394},
 primaryClass = {astro-ph.GA},
       adsurl = {https://ui.adsabs.harvard.edu/abs/2015ApJ...806..251J}
}

@ARTICLE{jjv2022,
       author = {{Jim{\'e}nez-Vicente}, J. and {Mediavilla}, E.},
        title = "{Fast Multipole Method for Gravitational Lensing: Application to High-magnification Quasar Microlensing}",
      journal = {\apj},
         year = 2022,
        month = dec,
       volume = {941},
       number = {1},
          eid = {80},
        pages = {80},
          doi = {10.3847/1538-4357/ac9e59},
archivePrefix = {arXiv},
       eprint = {2211.00354},
 primaryClass = {astro-ph.GA},
       adsurl = {https://ui.adsabs.harvard.edu/abs/2022ApJ...941...80J}
}

@ARTICLE{lemon2018,
       author = {{Lemon}, Cameron A. and {Auger}, Matthew W. and {McMahon}, Richard G. and {Ostrovski}, Fernanda},
        title = "{Gravitationally lensed quasars in Gaia - II. Discovery of 24 lensed quasars}",
      journal = {\mnras},
         year = 2018,
        month = oct,
       volume = {479},
       number = {4},
        pages = {5060-5074},
          doi = {10.1093/mnras/sty911},
archivePrefix = {arXiv},
       eprint = {1803.07601},
 primaryClass = {astro-ph.GA},
       adsurl = {https://ui.adsabs.harvard.edu/abs/2018MNRAS.479.5060L}
}

@ARTICLE{mchardy2018,
       author = {{McHardy}, I.~M. and {Connolly}, S.~D. and {Horne}, K. and {Cackett}, E.~M. and {Gelbord}, J. and {Peterson}, B.~M. and {Pahari}, M. and {Gehrels}, N. and {Goad}, M. and {Lira}, P. and {Arevalo}, P. and {Baldi}, R.~D. and {Brandt}, N. and {Breedt}, E. and {Chand}, H. and {Dewangan}, G. and {Done}, C. and {Elvis}, M. and {Emmanoulopoulos}, D. and {Fausnaugh}, M.~M. and {Kaspi}, S. and {Kochanek}, C.~S. and {Korista}, K. and {Papadakis}, I.~E. and {Rao}, A.~R. and {Uttley}, P. and {Vestergaard}, M. and {Ward}, M.~J.},
        title = "{X-ray/UV/optical variability of NGC 4593 with Swift: reprocessing of X-rays by an extended reprocessor}",
      journal = {\mnras},
         year = 2018,
        month = nov,
       volume = {480},
       number = {3},
        pages = {2881-2897},
          doi = {10.1093/mnras/sty1983},
archivePrefix = {arXiv},
       eprint = {1712.04852},
 primaryClass = {astro-ph.HE},
       adsurl = {https://ui.adsabs.harvard.edu/abs/2018MNRAS.480.2881M}
}

@ARTICLE{med2006,
       author = {{Mediavilla}, E. and {Mu{\~n}oz}, J.~A. and {Lopez}, P. and {Mediavilla}, T. and {Abajas}, C. and {Gonzalez-Morcillo}, C. and {Gil-Merino}, R.},
        title = "{A Fast and Very Accurate Approach to the Computation of Microlensing Magnification Patterns Based on Inverse Polygon Mapping}",
      journal = {\apj},
         year = 2006,
        month = dec,
       volume = {653},
       number = {2},
        pages = {942-953},
          doi = {10.1086/508796},
       adsurl = {https://ui.adsabs.harvard.edu/abs/2006ApJ...653..942M}
}

@ARTICLE{med2009,
       author = {{Mediavilla}, E. and {Mu{\~n}oz}, J.~A. and {Falco}, E. and {Motta}, V. and {Guerras}, E. and {Canovas}, H. and {Jean}, C. and {Oscoz}, A. and {Mosquera}, A.~M.},
        title = "{Microlensing-based Estimate of the Mass Fraction in Compact Objects in Lens Galaxies}",
      journal = {\apj},
         year = 2009,
        month = dec,
       volume = {706},
       number = {2},
        pages = {1451-1462},
          doi = {10.1088/0004-637X/706/2/1451},
archivePrefix = {arXiv},
       eprint = {0910.3645},
 primaryClass = {astro-ph.CO},
       adsurl = {https://ui.adsabs.harvard.edu/abs/2009ApJ...706.1451M}
}

@ARTICLE{med2011,
       author = {{Mediavilla}, E. and {Mediavilla}, T. and {Mu{\~n}oz}, J.~A. and {Ariza}, O. and {Lopez}, P. and {Gonzalez-Morcillo}, C. and {Jimenez-Vicente}, J.},
        title = "{New Developments on Inverse Polygon Mapping to Calculate Gravitational Lensing Magnification Maps: Optimized Computations}",
      journal = {\apj},
         year = 2011,
        month = nov,
       volume = {741},
       number = {1},
          eid = {42},
        pages = {42},
          doi = {10.1088/0004-637X/741/1/42},
       adsurl = {https://ui.adsabs.harvard.edu/abs/2011ApJ...741...42M}
}

@ARTICLE{med2011b,
       author = {{Mediavilla}, E. and {Mu{\~n}oz}, J.~A. and {Kochanek}, C.~S. and {Guerras}, E. and {Acosta-Pulido}, J. and {Falco}, E. and {Motta}, V. and {Arribas}, S. and {Manchado}, A. and {Mosquera}, A.},
        title = "{The Structure of the Accretion Disk in the Lensed Quasar SBS 0909+532}",
      journal = {\apj},
         year = 2011,
        month = mar,
       volume = {730},
       number = {1},
          eid = {16},
        pages = {16},
          doi = {10.1088/0004-637X/730/1/16},
       adsurl = {https://ui.adsabs.harvard.edu/abs/2011ApJ...730...16M}
}

@ARTICLE{med2025,
       author = {{Mediavilla}, E. and {Jim{\'e}nez-Vicente}, J. and {Heinze}, F.~M. and {Mart{\'\i}n Camalich}, J.},
        title = "{Quasar Milli-lensing and Cold Dark Matter (TNG50) Subhaloes. I. Image Splitting}",
      journal = {\apj},
         year = 2025,
        month = may,
       volume = {984},
       number = {2},
          eid = {145},
        pages = {145},
          doi = {10.3847/1538-4357/adc724},
       adsurl = {https://ui.adsabs.harvard.edu/abs/2025ApJ...984..145M}
}

@ARTICLE{miller2026,
       author = {{Miller}, Jake A. and {Cackett}, Edward M. and {Bentz}, Misty C. and {Goad}, Michael R. and {Korista}, Kirk T. and {McHardy}, Ian M. and {Carroll}, Russell W.},
        title = "{Continuum Reverberation Mapping of 18 Active Galactic Nuclei over 4 Yr}",
      journal = {\apj},
         year = 2026,
        month = jan,
       volume = {997},
       number = {1},
          eid = {62},
        pages = {62},
          doi = {10.3847/1538-4357/ae1c44},
archivePrefix = {arXiv},
       eprint = {2511.05669},
 primaryClass = {astro-ph.GA},
       adsurl = {https://ui.adsabs.harvard.edu/abs/2026ApJ...997...62M}
}

@ARTICLE{morgan2004,
       author = {{Morgan}, Nicholas D. and {Caldwell}, John A.~R. and {Schechter}, Paul L. and {Dressler}, Alan and {Egami}, Eiichi and {Rix}, Hans-Walter},
        title = "{WFI J2026-4536 and WFI J2033-4723: Two New Quadruple Gravitational Lenses}",
      journal = {\aj},
         year = 2004,
        month = may,
       volume = {127},
       number = {5},
        pages = {2617-2630},
          doi = {10.1086/383295},
archivePrefix = {arXiv},
       eprint = {astro-ph/0312478},
 primaryClass = {astro-ph},
       adsurl = {https://ui.adsabs.harvard.edu/abs/2004AJ....127.2617M}
}

@ARTICLE{morgan2010,
       author = {{Morgan}, Christopher W. and {Kochanek}, C.~S. and {Morgan}, Nicholas D. and {Falco}, Emilio E.},
        title = "{The Quasar Accretion Disk Size-Black Hole Mass Relation}",
      journal = {\apj},
         year = 2010,
        month = apr,
       volume = {712},
       number = {2},
        pages = {1129-1136},
          doi = {10.1088/0004-637X/712/2/1129},
archivePrefix = {arXiv},
       eprint = {1002.4160},
 primaryClass = {astro-ph.CO},
       adsurl = {https://ui.adsabs.harvard.edu/abs/2010ApJ...712.1129M}
}

@ARTICLE{motta2017,
       author = {{Motta}, V. and {Mediavilla}, E. and {Rojas}, K. and {Falco}, E.~E. and {Jim{\'e}nez-Vicente}, J. and {Mu{\~n}oz}, J.~A.},
        title = "{Probing the Broad-Line Region and the Accretion Disk in the Lensed Quasars HE 0435-1223, WFI 2033-4723, and HE 2149-2745 Using Gravitational Microlensing}",
      journal = {\apj},
         year = 2017,
        month = feb,
       volume = {835},
       number = {2},
          eid = {132},
        pages = {132},
          doi = {10.3847/1538-4357/835/2/132},
archivePrefix = {arXiv},
       eprint = {1703.00400},
 primaryClass = {astro-ph.GA},
       adsurl = {https://ui.adsabs.harvard.edu/abs/2017ApJ...835..132M}
}

@ARTICLE{netzer2025,
       author = {{Netzer}, Hagai},
        title = "{Disc winds spectral energy distribution and the variability time-scale of active galactic nuclei}",
      journal = {\mnras},
         year = 2025,
        month = jun,
       volume = {539},
       number = {4},
        pages = {3242-3249},
          doi = {10.1093/mnras/staf671},
archivePrefix = {arXiv},
       eprint = {2504.19344},
 primaryClass = {astro-ph.GA},
       adsurl = {https://ui.adsabs.harvard.edu/abs/2025MNRAS.539.3242N}
}

@ARTICLE{nierenberg2020,
       author = {{Nierenberg}, A.~M. and {Gilman}, D. and {Treu}, T. and {Brammer}, G. and {Birrer}, S. and {Moustakas}, L. and {Agnello}, A. and {Anguita}, T. and {Fassnacht}, C.~D. and {Motta}, V. and {Peter}, A.~H.~G. and {Sluse}, D.},
        title = "{Double dark matter vision: twice the number of compact-source lenses with narrow-line lensing and the WFC3 grism}",
      journal = {\mnras},
         year = 2020,
        month = mar,
       volume = {492},
       number = {4},
        pages = {5314-5335},
          doi = {10.1093/mnras/stz3588},
archivePrefix = {arXiv},
       eprint = {1908.06344},
 primaryClass = {astro-ph.GA},
       adsurl = {https://ui.adsabs.harvard.edu/abs/2020MNRAS.492.5314N}
}

@ARTICLE{mudd2018,
       author = {{Mudd}, D. and {Martini}, P. and {Zu}, Y. and {Kochanek}, C. and {Peterson}, B.~M. and {Kessler}, R. and {Davis}, T.~M. and {Hoormann}, J.~K. and {King}, A. and {Lidman}, C. and {Sommer}, N.~E. and {Tucker}, B.~E. and {Asorey}, J. and {Hinton}, S. and {Glazebrook}, K. and {Kuehn}, K. and {Lewis}, G. and {Macaulay}, E. and {Moeller}, A. and {O'Neill}, C. and {Zhang}, B. and {Abbott}, T.~M.~C. and {Abdalla}, F.~B. and {Allam}, S. and {Banerji}, M. and {Benoit-L{\'e}vy}, A. and {Bertin}, E. and {Brooks}, D. and {Carnero Rosell}, A. and {Carollo}, D. and {Carrasco Kind}, M. and {Carretero}, J. and {Cunha}, C.~E. and {D'Andrea}, C.~B. and {da Costa}, L.~N. and {Davis}, C. and {Desai}, S. and {Doel}, P. and {Fosalba}, P. and {Garc{\'\i}a-Bellido}, J. and {Gaztanaga}, E. and {Gerdes}, D.~W. and {Gruen}, D. and {Gruendl}, R.~A. and {Gschwend}, J. and {Gutierrez}, G. and {Hartley}, W.~G. and {Honscheid}, K. and {James}, D.~J. and {Kuhlmann}, S. and {Kuropatkin}, N. and {Lima}, M. and {Maia}, M.~A.~G. and {Marshall}, J.~L. and {McMahon}, R.~G. and {Menanteau}, F. and {Miquel}, R. and {Plazas}, A.~A. and {Romer}, A.~K. and {Sanchez}, E. and {Schindler}, R. and {Schubnell}, M. and {Smith}, M. and {Smith}, R.~C. and {Soares-Santos}, M. and {Sobreira}, F. and {Suchyta}, E. and {Swanson}, M.~E.~C. and {Tarle}, G. and {Thomas}, D. and {Tucker}, D.~L. and {Walker}, A.~R. and {DES Collaboration}},
        title = "{Quasar Accretion Disk Sizes from Continuum Reverberation Mapping from the Dark Energy Survey}",
      journal = {\apj},
         year = 2018,
        month = aug,
       volume = {862},
       number = {2},
          eid = {123},
        pages = {123},
          doi = {10.3847/1538-4357/aac9bb},
archivePrefix = {arXiv},
       eprint = {1711.11588},
 primaryClass = {astro-ph.GA},
       adsurl = {https://ui.adsabs.harvard.edu/abs/2018ApJ...862..123M}
}

@ARTICLE{oguri2008d,
       author = {{Oguri}, Masamune and {Inada}, Naohisa and {Blackburne}, Jeffrey A. and {Shin}, Min-Su and {Kayo}, Issha and {Strauss}, Michael A. and {Schneider}, Donald P. and {York}, Donald G.},
        title = "{Mass models and environment of the new quadruply lensed quasar SDSS J1330+1810}",
      journal = {\mnras},
         year = 2008,
        month = dec,
       volume = {391},
       number = {4},
        pages = {1973-1980},
          doi = {10.1111/j.1365-2966.2008.14032.x},
archivePrefix = {arXiv},
       eprint = {0809.0913},
 primaryClass = {astro-ph},
       adsurl = {https://ui.adsabs.harvard.edu/abs/2008MNRAS.391.1973O}
}

@BOOK{peterson1997,
       author = {{Peterson}, Bradley M.},
        title = "{An Introduction to Active Galactic Nuclei}",
         year = 1997,
       adsurl = {https://ui.adsabs.harvard.edu/abs/1997iagn.book.....P}
}

@ARTICLE{ss1973,
       author = {{Shakura}, N.~I. and {Sunyaev}, R.~A.},
        title = "{Black holes in binary systems. Observational appearance.}",
      journal = {\aap},
         year = 1973,
        month = jan,
       volume = {24},
        pages = {337-355},
       adsurl = {https://ui.adsabs.harvard.edu/abs/1973A&A....24..337S}
}

\clearpage

\begin{appendix}

\onecolumn
\section{Gravitationally lensed quasar information for simulated magnification maps}

The convergence ($\kappa$) and external shear ($\gamma$) used for the simulated magnification maps in Section \ref{sec:method} is obtained from the literature and presented in Table \ref{tab:kg}.

\begin{table*}[htb!]
\caption{Convergence and Shear}  
\label{tab:kg} 
\centering 
\begin{tabular}{l c c c c c l} 
\hline\hline  
System & Image & $z_L$ & $z_S$ & $\kappa$ & $\gamma$ & Reference\tablefootmark{d} \\   \hline    
DESJ0405-3308 & A & 0.5\tablefootmark{a} & 1.713 & 0.48 & 0.46 &  \cite{shajib2019};  \\	
              & B &   & & 0.53 & 0.53 & \cite{anguita2018}\\
              & C &   & & 0.52 & 0.51 & \\  
              & D &   & & 0.49 & 0.47 & \\ \hline
RXJ0911+0551  & A &   0.77 & 2.763 & 0.646 & 0.544 & \cite{schechter2014}; \\	
              & B &   & & 0.586 & 0.281 &  \cite{bade1997} \\
              & C &   & & 0.637 & 0.577 & \\  
              & D &   & & 0.290 & 0.066 & \\ \hline
SDSSJ1330+1810 & A & 0.373 & 1.383 & 0.48 & 0.43 & \cite{shajib2019};  \\	
              & B &   & & 0.59 & 0.52 & \cite{oguri2008d}\\
              & C &   & & 0.36 & 0.43 & \\  
              & D &   & & 0.74 & 0.71 & \\ \hline
PSJ1606-2333 & A & 0.5\tablefootmark{a} & 1.696 & 0.46 & 0.25 & \cite{shajib2019};\\	
             & B &   & & 0.49 & 0.22 &  \cite{lemon2018}\\
             & C &   & & 0.77 & 0.75 & \\  
             & D &   & & 0.57 & 0.66 & \\ \hline
WFI2026-4536 & A$_1$ & 1.04\tablefootmark{b} & 2.23 & 0.52 & 0.41 & \cite{cornachione2020}\tablefootmark{c};  \\	
             & A$_2$ &   & & 0.53 & 0.53 & \cite{morgan2004}\\
             & B     &   & & 0.43 & 0.32 & \\  
             & C     &   & & 0.53 & 0.65 & \\ \hline
WFI2033-4723 & A$_1$ & 0.66 & 1.66 & 0.506 & 0.255 & \cite{schechter2014}; \\	
             & A$_2$ &   & & 0.665 & 0.643 & \cite{morgan2004}\\
             & B     &   & & 0.392 & 0.302 & \\  
             & C     &   & & 0.700 & 0.735 & \\ \hline
DESJ2038-4008 & A & 0.228 & 0.777 & 0.21 & 0.43 & \cite{shajib2019}; \\	
              & B &   & & 0.22 & 0.49 & \cite{agnello2018a}\\
              & C &   & & 0.45 & 0.89 & \\  
              & D &   & & 0.59 & 1.11 & \\ \hline  
\end{tabular}
\tablefoot{} 
\tablefoottext{a}{Unknown redshift. \cite{shajib2019} assumed this value to estimate $\kappa$ and $\gamma$.}
\tablefoottext{b}{Unconfirmed spectroscopic redshift. \cite{cornachione2020} indicates this is the value compatible with featurless spectra and time delay measurements.}
\tablefoottext{c}{Values obtained from \cite{cornachione2020} to produce $\kappa_{\star}/\kappa\sim 0.1$ (i.e., lens model corresponding to $f_M/f_L=0.3$).}
\tablefoottext{d}{Reference for deflector and source redshifts.}
\end{table*}

\section{Simulated individual probability distributions of microlensing maps \label{sec:maps}}

We generate microlensing magnification maps for each lensed image using the inverse polygon mapping algorithm \citep{med2006,med2011,jjv2022}.
We assume that 10\% of the surface mass density correspond to stars. This is a reasonable choice for quads, where the images are typically formed at distances greater than $\sim1.5$ effective radii. 
This value of the stellar mass fraction may fall in the lower range of the estimates by  \cite{jjv2015}; however, it represents a conservative choice in the context of our study, as a low mass fraction would bias the sizes toward smaller values (i.e., favoring convergence with the thin disk model.) We adopt masses\footnote{In Figure \ref{fig:salpeter} we show the negligible impact of considering a mass function for the stars.} $M=1 M_{\odot}$. Note that, owing to the mass-size degeneracy, the estimated sizes ($r_s$) can be rescaled to other masses ($M$) as $r_s(\lambda) \propto (M/M_{\odot})^{1/2}$.
The values for the convergence ($\kappa$) and shear ($\gamma$) are obtained from the literature (see Table \ref{tab:kg}). 
The map sizes range from $1300\times1300$ squared pixels up to $1900\times1900$ squared pixels, with the same pixel size of $\sim0.4$~lt-days. For the objects in our sample this size corresponds to 30 Einstein radii. To improve the statistics in those cases with a small number of stars, where sample variance may be important, we build five\footnote{We arbitrarily started with five realizations; since the final results differ by less than 1 light-day, increasing their number is unnecessary.} microlensing maps for each image, yielding five combined estimations for $r_s$ and $p$. Hereafter, our results refers to the average of those five maps.

\begin{figure*}[htb!]
        \centering      
        \includegraphics[width=4.7cm]{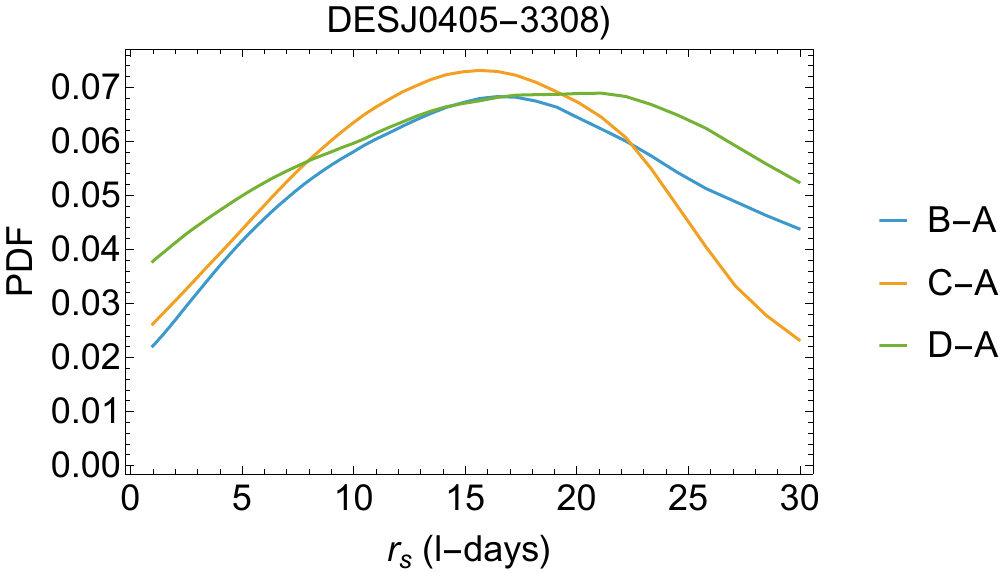}
        \includegraphics[width=4.0cm]{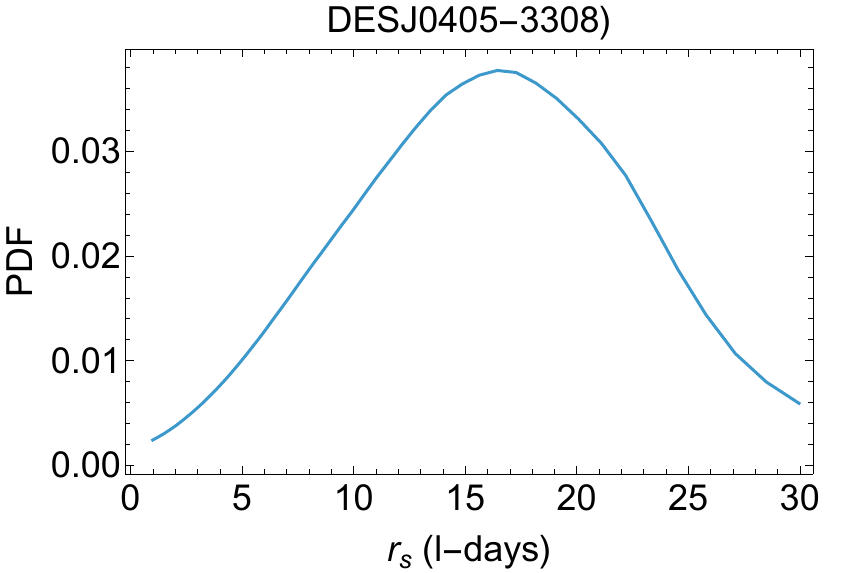}
        \includegraphics[width=4.7cm]{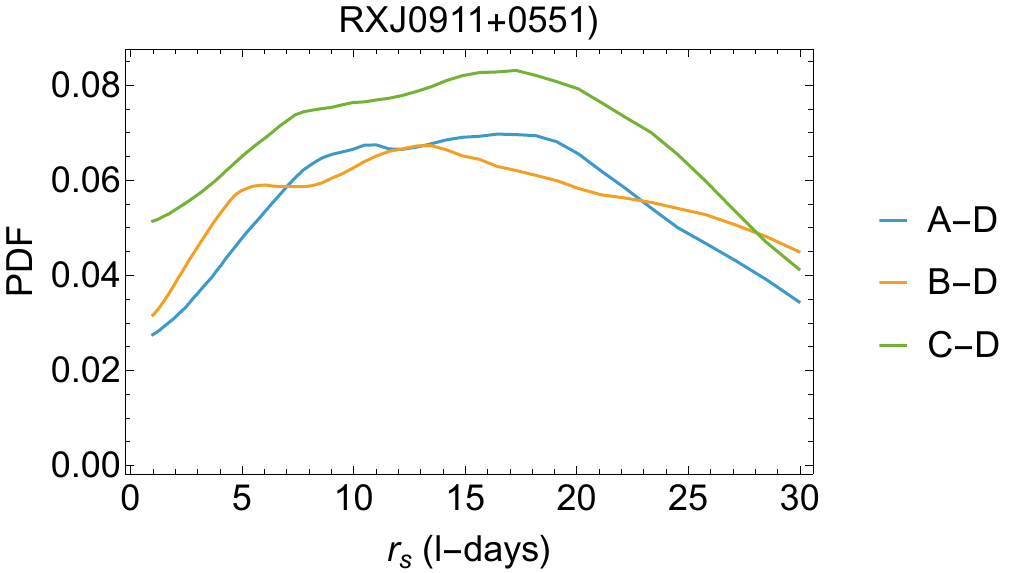}
        \includegraphics[width=4.0cm]{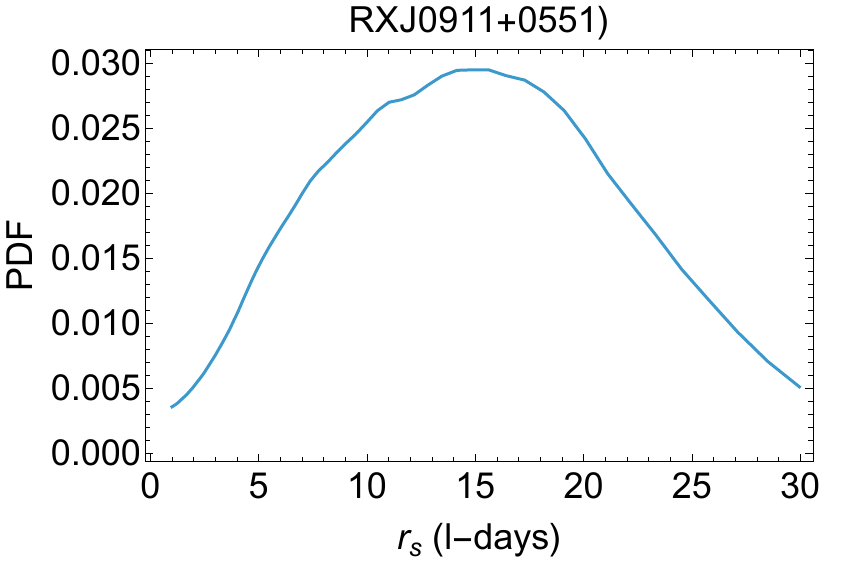}\\
        \includegraphics[width=4.7cm]{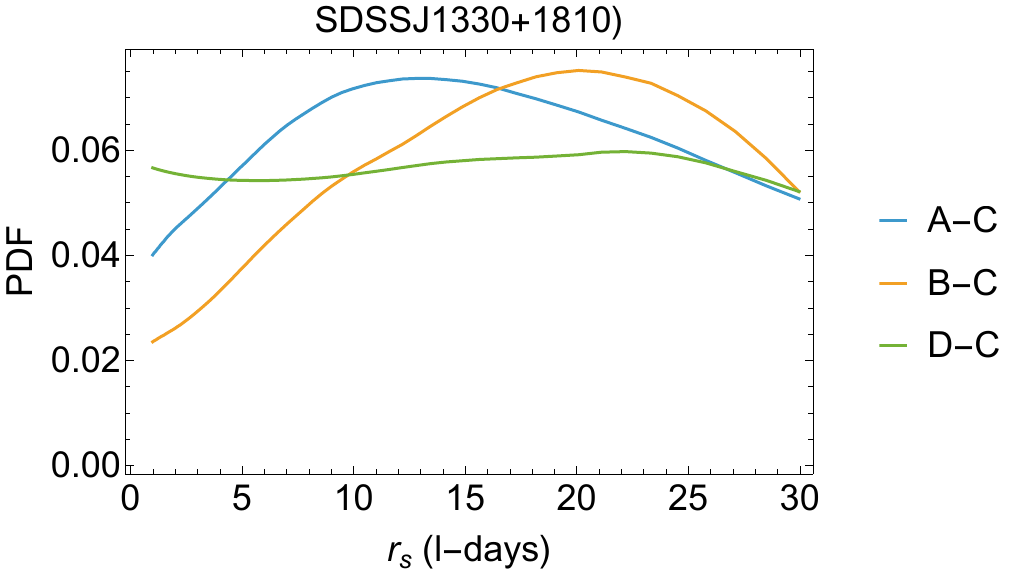}
        \includegraphics[width=4.0cm]{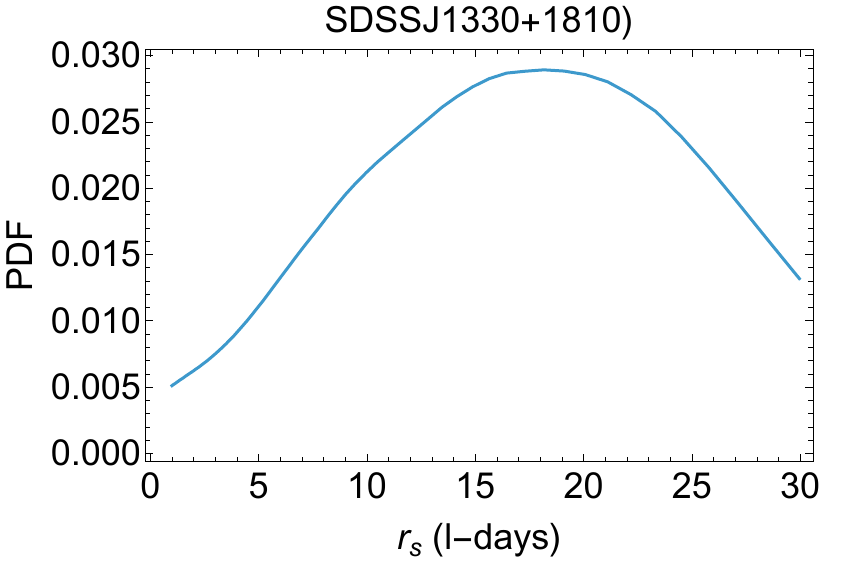}
        \includegraphics[width=4.7cm]{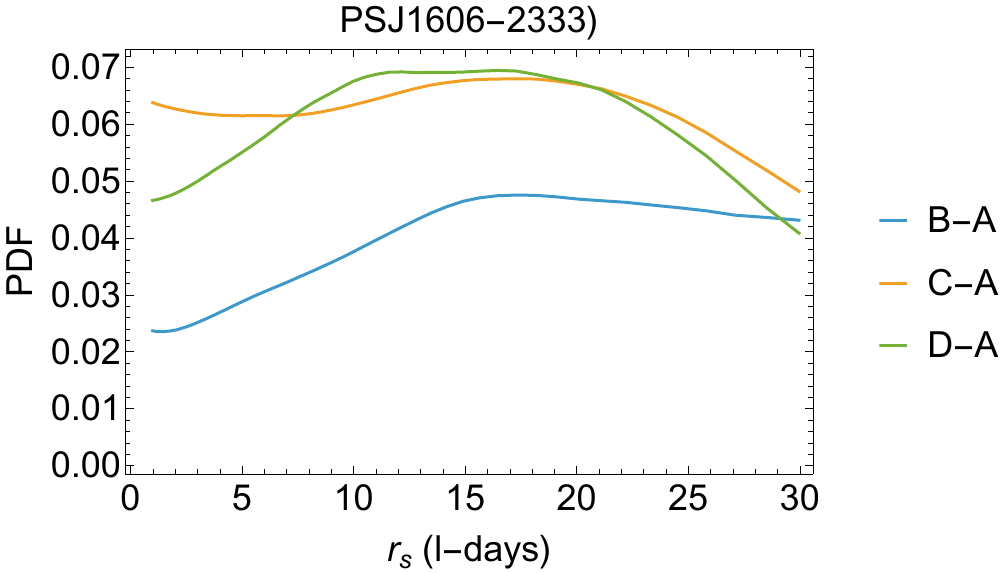}
        \includegraphics[width=4.0cm]{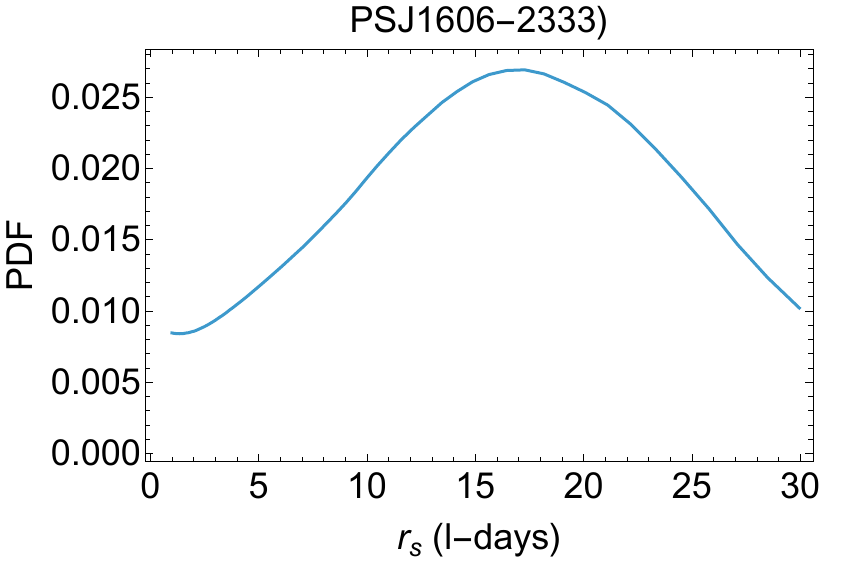}\\
        \centering      
        \includegraphics[width=4.7cm]{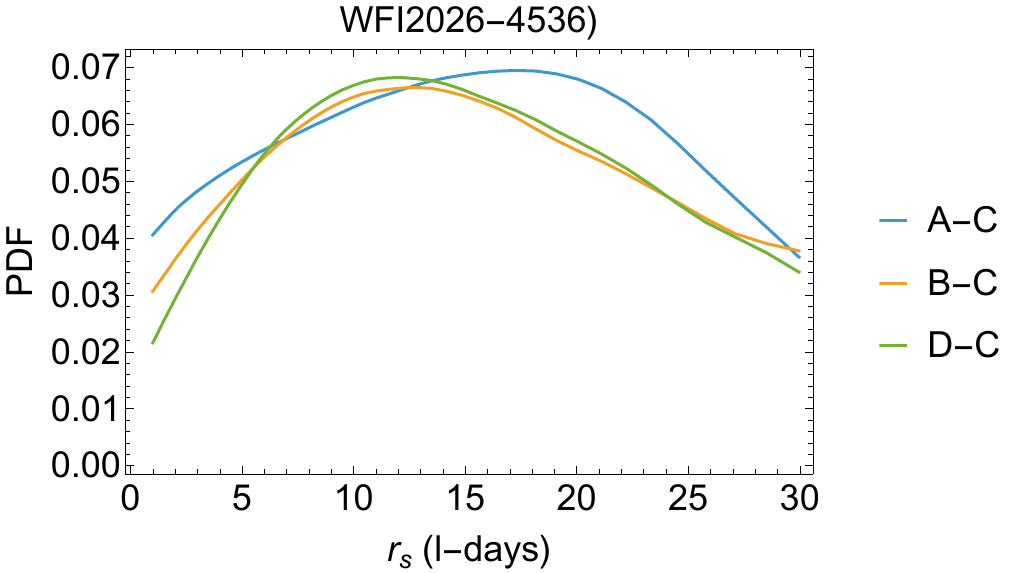}
        \includegraphics[width=4.0cm]{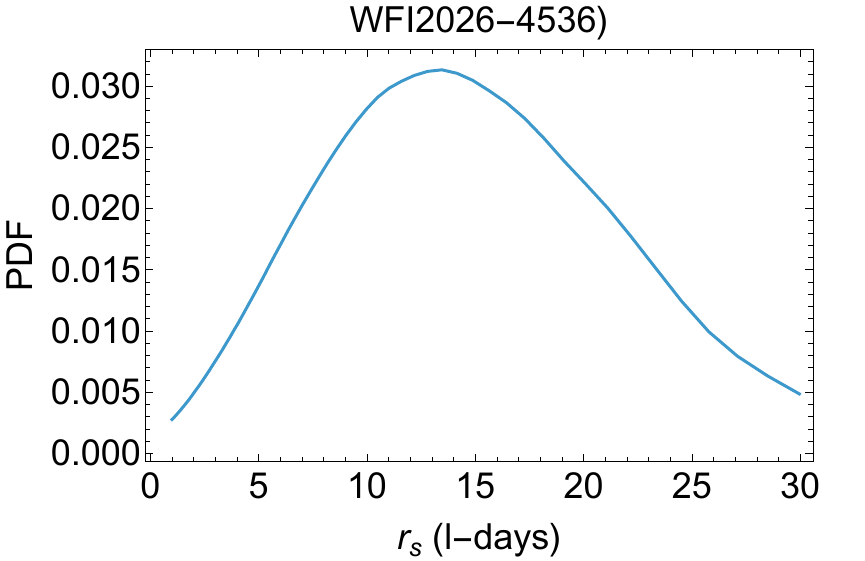} 
        \includegraphics[width=4.7cm]{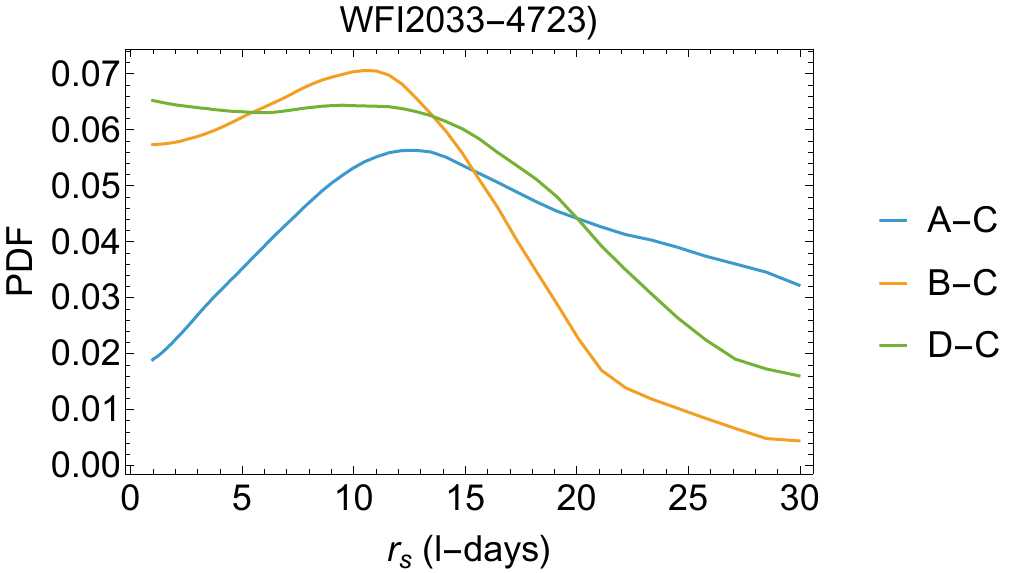}
        \includegraphics[width=4.0cm]{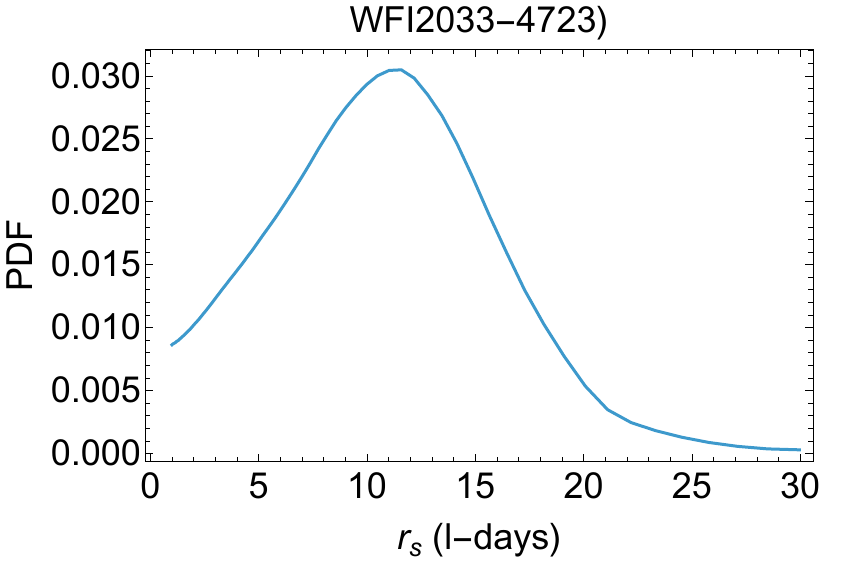}\\    
        \includegraphics[width=4.7cm]{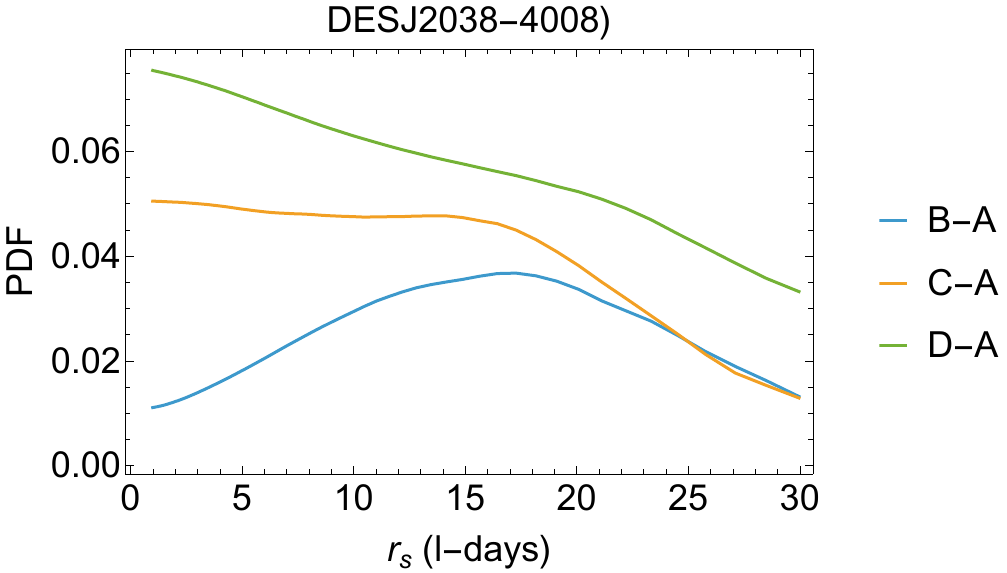}
        \includegraphics[width=4.0cm]{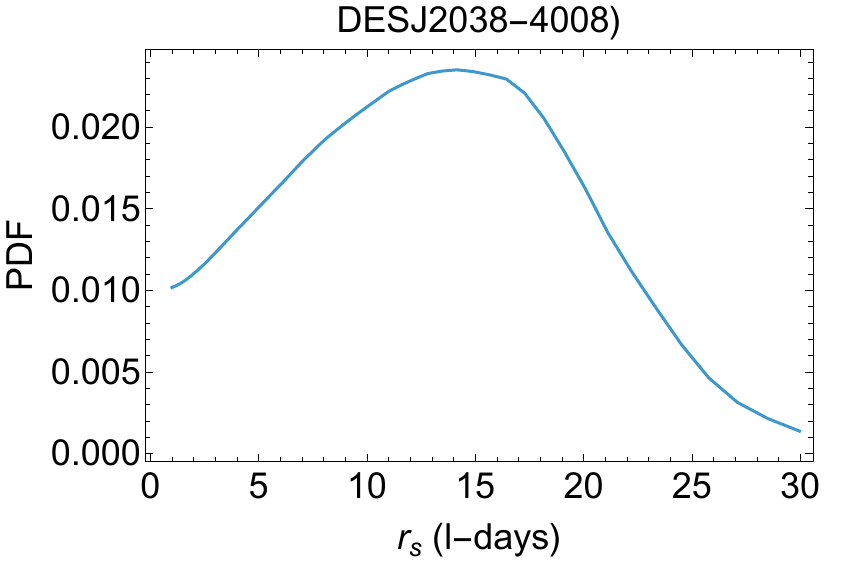}\\    
      \caption{Individual and conflated PDFs for each system in the last iteration for one realization of microlensing maps. }
\label{fig:pdfs}
    \end{figure*}

\begin{figure*}[htb!]
        \centering      
        \includegraphics[width=10cm]{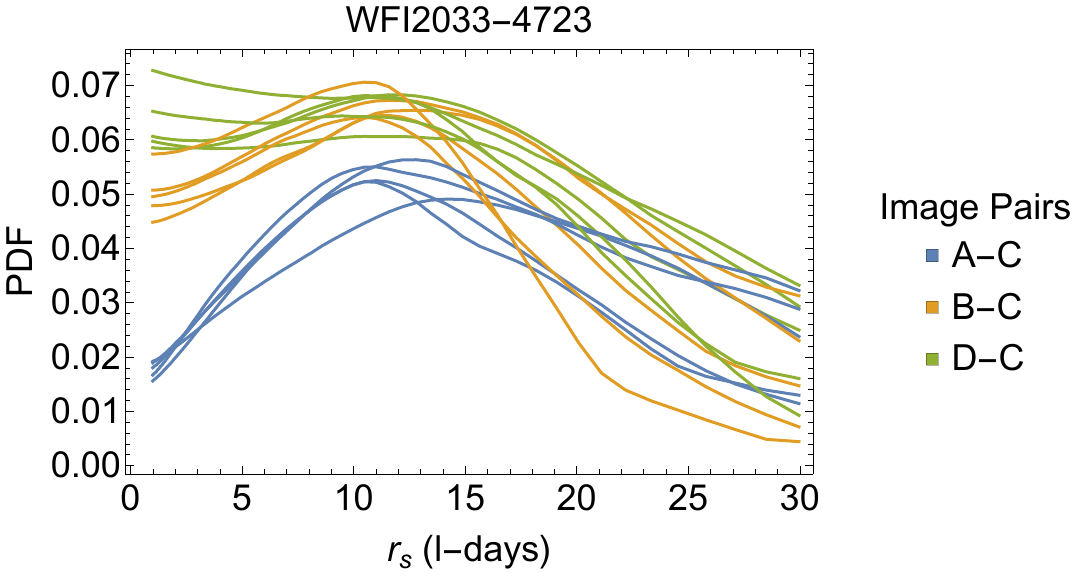}
        \includegraphics[width=8cm]{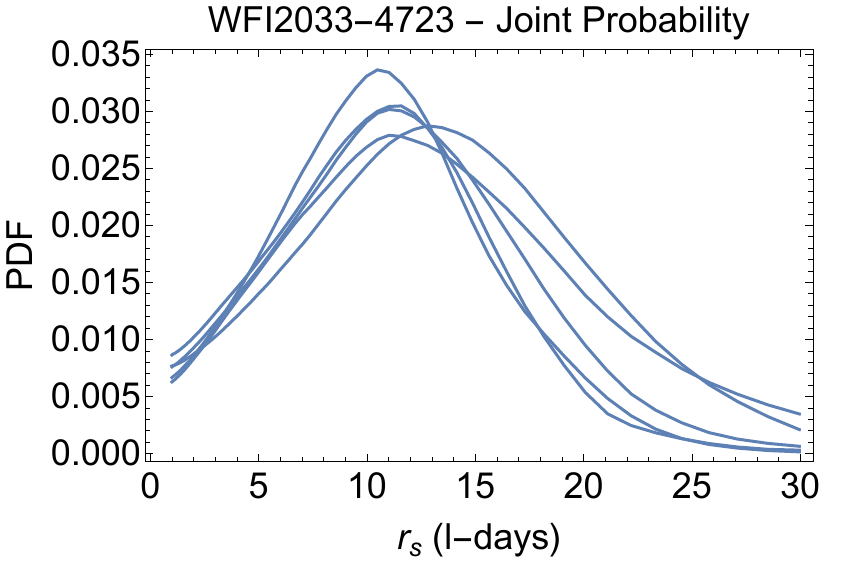}\\    
\caption{PDFs for WFI2033-4723 for the five realizations. Left panel: individual PDFs for the pairs AC (blue), BC (orange), and DC (green). Right panel: conflated PDFs. }
         \label{fig:pdfs2}
    \end{figure*}

The measured microlensing magnifications are shown in Table \ref{tab:dm}. The individual PDF for each pair of images as well as the conflated PDF for each system is presented in Figure \ref{fig:pdfs}.  As an example, Figure \ref{fig:pdfs2} shows the individual and conflated PDF for each realization for WFI2033-4723.

\begin{table}[htb!]
\caption{Microlensing magnification amplitude}  
\label{tab:dm} 
\centering 
\begin{tabular}{l c c r} 
\hline\hline  
System & line (\AA) & pair & $\Delta m^{obs}_i$ (mag) \\    
\hline                 
DESJ0405-3308 & [OIII] & BA & $-0.13 \pm 0.03$ \\	
              &      & CA & $0.39 \pm 0.06$ \\
              &      & DA & $0.05 \pm 0.06$ \\  \hline
RXJ0911+0551  & [NeIII] & AD & $0.56 \pm0.06$ \\ 
              &      & BD & $0.63 \pm 0.17$ \\
              &      & CD & $0.76 \pm 0.18$ \\ \hline
SDSSJ1330+1810 & [NeIII] & AC & $0.10 \pm 0.08$ \\ 
              &      & BC & $0.55 \pm 0.13$ \\
              &      & DC & $0.99 \pm 0.16$ \\ \hline
PSJ1606-2333 & [OIII] & BA & $-0.15 \pm 0.04$ \\  
             &       & CA & $-0.18 \pm 0.04$ \\ 
             &       & DA & $0.09 \pm 0.04$ \\ \hline
WFI2026-4536 & [OIII] & A$_1$B & $-0.11 \pm 0.06$ \\ 
             &         & A$_2$B & $0.05 \pm 0.08$ \\
             &         & CB & $0.15 \pm 0.09$ \\ \hline
WFI2033-4723 & [OIII] & A$_1$B & $0.10 \pm 0.04$ \\ 
             &         & A$_2$B & $0.25 \pm 0.07$ \\	
             &         & CB & $0.33 \pm 0.06$ \\ \hline
DESJ2038-4008 & [OIII] & AB & $-0.23 \pm 0.17$ \\
             &         & CB & $-0.41 \pm 0.12$ \\
             &         & DB & $-0.20 \pm 0.12$ \\ \hline  
\end{tabular}
\tablefoot{$\Delta m^{obs}_i$ is the observed magnitude difference between the NEL and its underlying continuum for each pair of lensed images.}
\end{table}

\section{Accretion disk size and temperature profile estimation from microlensing measurements}

\begin{figure*}[htb!]
        \centering      
        \includegraphics[width=9cm]{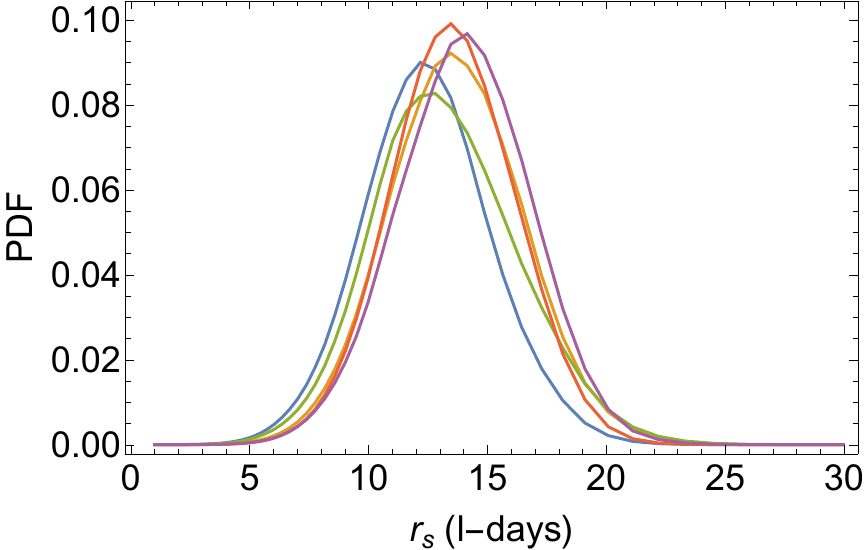}
      \includegraphics[width=9cm]{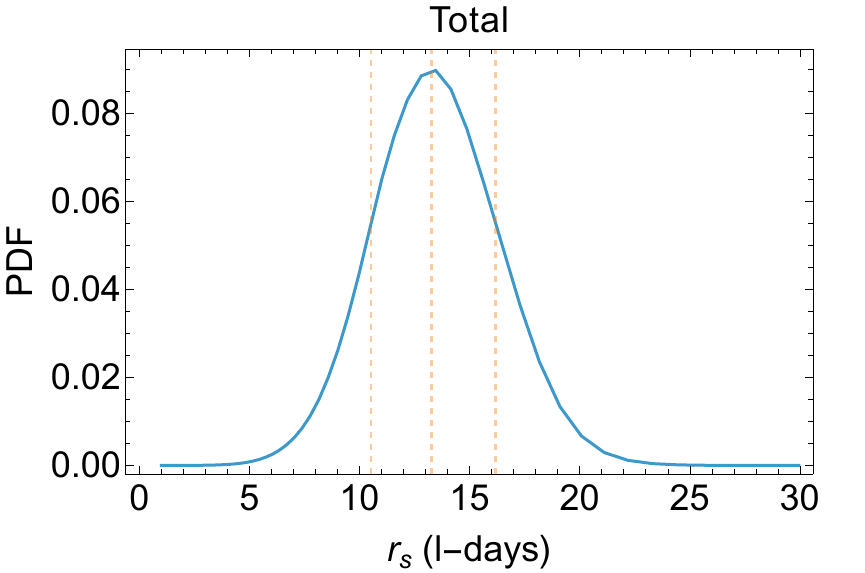}
       \caption{Left panel: joint PDFs for $r_s$ at $\lambda_{ref}=5007$\AA\, for the five realizations of microlensing maps in the final iteration.  Right panel: joint PDF from the individual PDFs (see Tab. \ref{tab:rsvalues}). }
        \label{fig:jointpdf}
   \end{figure*}

The temperature slope ($p$) is obtained from an iterative process described in Section \ref{sec:method}. Table \ref{tab:pvalues} shows our results for $r_s$ and $p$ together with the literature data used in Figure \ref{fig:p}. The joint PDFs for the last iteration is shown in the left panel of Figure \ref{fig:jointpdf}. Size estimation for each individual system are shown in Table \ref{tab:rsvalues}, and their joint PDF is presented in the right panel of Figure \ref{fig:jointpdf}.

\begin{table}[htb!]
\caption{Accretion disk size and temperature profile}   
\label{tab:pvalues}    
\centering                       
\begin{tabular}{c c c c c}    
\hline\hline               
$\lambda_{ref}$ (\AA) & Run & $r_s$ (l-d) & $p$ &  scale\tablefootmark{a}  \\         
\hline                      
10\tablefootmark{b} & - &  $1.3_{-0.65}^{+1.3}$ & - & - \\   
10\tablefootmark{c} & - &  $1.4\pm2.1$ & - & - \\   
\hline
1026\tablefootmark{d} & - &  $4.5_{-1.2}^{+1.5}$ & $0.8\pm0.2$ & - \\ 
$2500$\tablefootmark{e} & - & $5.7^{+1.8}_{-1.4}$ & - & - \\
$2500$\tablefootmark{f} & - & $7.7_{-1.5}^{+2.2}$ & - & \\
\hline
5007 & 511 & $12.8_{-2.7}^{+2.9}$ & $0.66\pm0.23$ & $1.33$ \\
     & 512 & $14.1_{-2.9}^{+3.0}$ & $0.72\pm0.23$ & $1.33$ \\
     & 513 & $13.7_{-2.9}^{+3.3}$ & $0.70\pm0.24$ & $1.33$ \\
     & 514 & $13.9_{-2.7}^{+2.8}$ & $0.71\pm0.23$ & $1.33$ \\
     & 515 & $14.3_{-2.9}^{+2.9}$ & $0.73\pm0.23$ & $1.33$ \\
     \hline
5007 & 521 & $12.4_{-2.6}^{+2.7}$ & $0.64\pm0.23$ & $0.66$ \\
     & 522 & $13.7_{-2.8}^{+2.9}$ & $0.70\pm0.23$ & $0.72$ \\
     & 523 & $13.1_{-2.8}^{+3.3}$ & $0.68\pm0.24$ & $0.69$ \\
     & 524 & $13.5_{-2.6}^{+2.7}$ & $0.69\pm0.23$ & $0.69$ \\
     & 525 & $14.0_{-2.8}^{+2.8}$ & $0.72\pm0.23$ & $0.72$ \\
     \hline
5007 & 531 & $12.4_{-2.6}^{+2.7}$ & $0.64\pm0.23$ & $0.63$ \\
     & 532 & $13.6_{-2.8}^{+2.9}$ & $0.70\pm0.23$ & $0.69$ \\
     & 533 & $13.1_{-2.8}^{+3.3}$ & $0.67\pm0.24$ & $0.66$ \\
     & 534 & $13.5_{-2.6}^{+2.7}$ & $0.69\pm0.23$ & $0.66$ \\
     & 535 & $14.0_{-2.8}^{+2.8}$ & $0.72\pm0.23$ & $0.72$ \\
\hline  
5007 & 541 & $12.4_{-2.6}^{+2.7}$ & $0.64\pm0.23$ & $0.63$ \\
     & 542 & $13.7_{-2.8}^{+2.9}$ & $0.70\pm0.23$ & $0.69$ \\
     & 543 & $13.1_{-2.8}^{+3.3}$ & $0.67\pm0.24$ & $0.66$ \\
     & 544 & $13.5_{-2.6}^{+2.7}$ & $0.69\pm0.23$ & $0.69$ \\
     & 545 & $14.0_{-2.8}^{+2.8}$ & $0.72\pm0.23$ & $0.72$ \\
\hline  
\end{tabular}
\tablefoot{} 
\tablefoottext{a}{Initial value for $p$ used to scale [NeIII]$\lambda3969$ to [OIII]$\lambda5007$.}
\tablefoottext{b}{Values estimated from a sample of 30 image-pairs in X-ray and 18 image-pairs in optical by \cite{jjv2015}.}
\tablefoottext{b}{Average value estimated from 10 systems in X-rays from  \cite{pooley2007}.}
\tablefoottext{d}{Values estimated from 10 image pairs with optical spectra by \cite{jjv2014}.}
\tablefoottext{e}{Value estimated from equation 4 for $10^9 \rm M_{\odot}$ \citep{morgan2010}. Notice that the radius obtained from that equation should be multiplied by 2.44 to obtain $r_s$.}
\tablefoottext{f}{Estimated value from figure 10 for $10^9 \rm M_{\odot}$ \citep{black2011}.}
\end{table}

\begin{table}[htb!]
\caption{Individual accretion disk size estimation at reference wavelength $\lambda 5007 \AA$.}              
\label{tab:rsvalues}    
\centering                        
\begin{tabular}{l c }      
\hline\hline               
Object &  $r_s$   \\
       & (l-days)  \\
\hline                      
 \text{DESJ0405-3308} & $15.0_{6.6}^{+7.3}$ \\
 \text{RXJ0911+0551}    & $13.9_{6.4}^{+7.1}$ \\
 \text{SDSSJ1330+1810}  & $16.8_{7.9}^{+7.8}$ \\
 \text{PSJ1606-2333}    & $16.0_{7.8}^{+7.7}$ \\
 \text{WFI2026-4536}    & $13.5_{6.0}^{+7.3}$ \\
 \text{WFI2033-4723}    & $11.8_{5.6}^{+6.3}$ \\
 \text{DESJ2038-4008} & $13.3_{7.3}^{+7.2}$ \\
\hline
\end{tabular}
\end{table}

\section{Comparison of magnification maps obtained from monocromatic and Salpeter mass distributions}

\begin{figure*}[htb!]
	\centering      
  	\includegraphics[width=8cm]{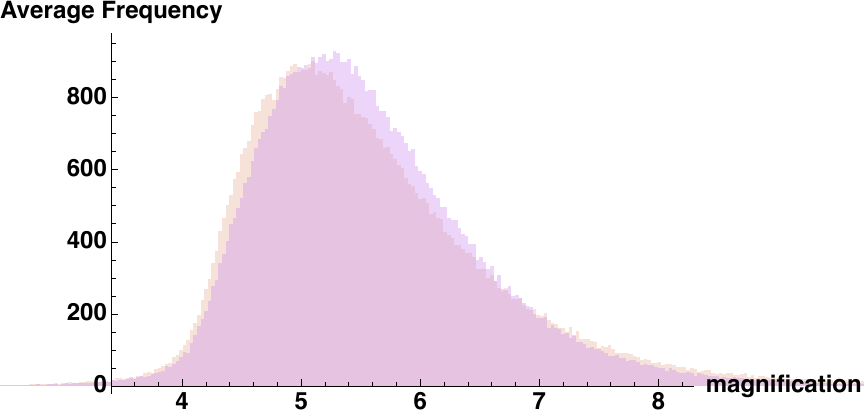} 
	\includegraphics[width=8cm]{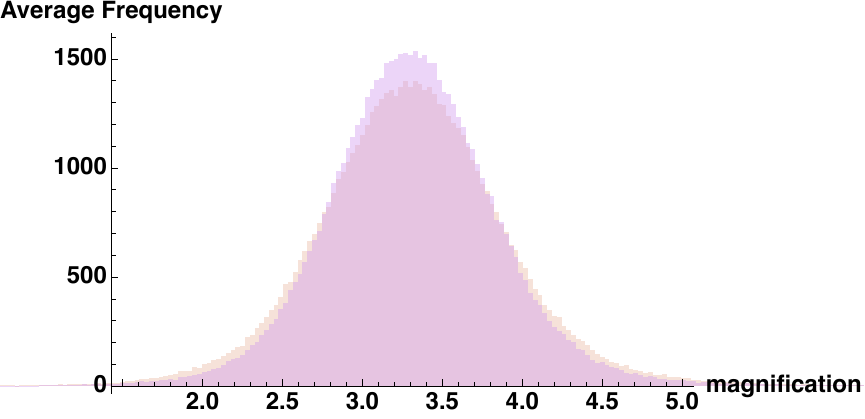}
	\includegraphics[width=8cm]{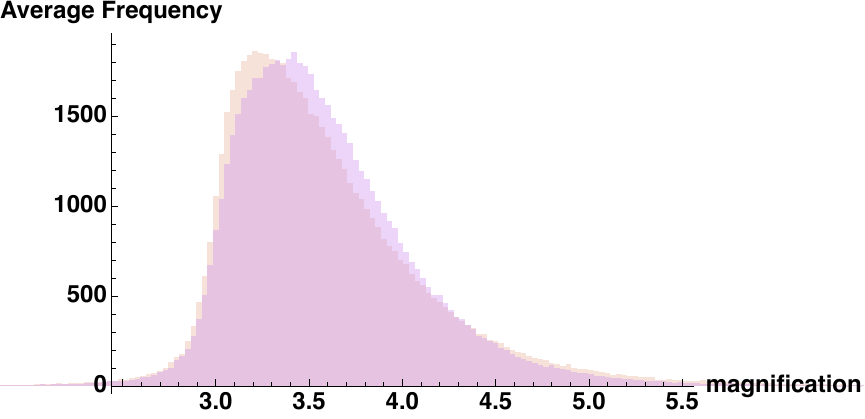}
	\includegraphics[width=8cm]{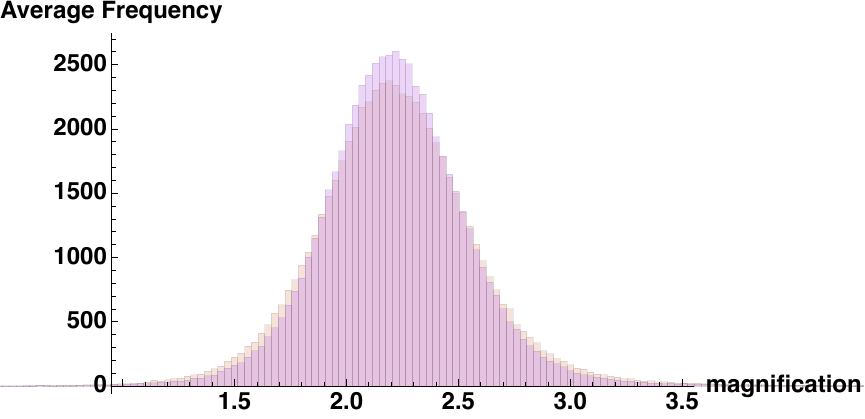}  
	\caption{Comparison of the magnification histograms obtained from different stellar mass distributions for WFI2033-4723 convolved with the size of the source $r_s=13.3$ lt-days. Purple corresponds to the Salpeter mass distribution between 0.2 $M_\odot$ and 1.0 $M_\odot$ while orange corresponds to the monochromatic mass distribution for $0.35M_\odot$ (the average mass of the Salpeter distribution). From top to bottom, left to right: combination of the five realization of microlensing maps for A, B, C, and D.} 
	\label{fig:salpeter}
\end{figure*}

The shape of the microlenses mass function is only
expected to be important for markedly bimodal distributions (Schechter et al. 2014) 
with a large and comparable contribution to the mass density from microlenses of very different masses. Otherwise, for smooth mass functions that span a relatively narrow range of masses, it is usually assumed that the only relevant parameter is the average mass, implicitly taken as the arithmetic mean, although the scale-invariant geometric mean may be preferable (Esteban-Gutierrez et al. 2020).  Nevertheless, we have reproduced our calculations for WFI2033-4723 to assess the impact of such assumption. We used a Salpeter mass distribution ($\alpha=2.3$) with a mass range $0.2-1.0 M_{\odot}$.  
The magnification maps resolution is 1.6~lt-days and their size $500\times500$ pixels. Given that the geometrical mean mass of the Salpeter distribution is $M_{micro}=0.35 M_{\odot}$, we also reproduced the same  maps with a monochromatic mass distribution of this value. Each map is convolved with a source size $r_x=13.3$~lt-days.  Figure \ref{fig:salpeter} shows the comparison of the microlensing magnification histograms for the four images combining the five realizations. The histograms are very similar.

\end{appendix}

\end{document}